\documentclass[aps,manuscript,showpacs,showkeys,superscriptaddress,nofootinbib]{revtex4-1}
\usepackage{graphicx}
\usepackage{epsfig}  
\usepackage{epsf}    
\usepackage{dcolumn}
\usepackage{bm}
\usepackage{dcolumn}
\usepackage{textcomp}
\usepackage[tbtags]{amsmath}
\usepackage{amsfonts}
\usepackage{float}
\usepackage{subfig}
\usepackage[]{hyperref}
  \hypersetup{
  unicode=false,          
  pdftoolbar=true,        
  pdfmenubar=true,        
  pdffitwindow=true,     
  pdfstartview={FitH},    
  pdfsubject={Getting the best out of T2K and NOvA},   
  pdfnewwindow=true,      
  pdfcreator={RevTeX},
  colorlinks=true,       
  linkcolor=red,          
  citecolor=blue,        
  urlcolor=blue,           
  }
\usepackage{hypcap}

\def\anu{{\bar\nu}}

\newcommand{\beq}{\begin{equation}}
\newcommand{\eeq}{\end{equation}}
\newcommand{\beqa}{\begin{eqnarray}}
\newcommand{\eeqa}{\end{eqnarray}}

\newcommand{\tx}{{\theta_{12}}}
\newcommand{\ty}{{\theta_{13}}}
\newcommand{\tz}{{\theta_{23}}}

\newcommand{\dl}{{\Delta_{31}}}
\newcommand{\ds}{{\Delta_{21}}}

\newcommand{\ahat}{\hat{A}}
\newcommand{\dhat}{\hat{\Delta}}

\newcommand{\dcp}{\delta_{\mathrm{CP}}}
\newcommand{\nova}{NO$\nu$A~}
\newcommand{\pmue}{P(\nu_\mu \rightarrow \nu_e)}
\newcommand{\pmumu}{P(\nu_\mu \rightarrow \nu_\mu)}
\newcommand{\pme}{P_{\mu e}}
\newcommand{\pmuebar}{P(\bar{\nu}_{\mu} \rightarrow \bar{\nu}_e)}
\newcommand{\pmumubar}{P(\bar{\nu}_{\mu} \rightarrow \bar{\nu}_\mu)}
\newcommand{\pmebar}{P_{\bar{\mu} \bar{e}}}

\newcommand{\dchsq}{\Delta\chi^2}

\begin{document}


\title{On the tension between the latest \nova and T2K data}

\author{Ushak Rahaman}
\email[Email Address: ]{ushakr@uj.ac.za}
\affiliation{Centre for Astro-Particle Physics (CAPP) and Department of Physics, University of Johannesburg, PO Box 524, Auckland Park 2006, South Africa}
\author{Sushant K. Raut}
\email[Email Address: ]{sushant.raut@krea.edu.in}
\affiliation{Division of Sciences, Krea University, Sri City, India 517646}

\date{\today}
\begin{abstract}

\end{abstract}
\begin{abstract}
The latest data from the T2K and \nova experiments show a tension in their preferred values of the oscillation parameters. In this work, we try to identify the source of the tension between the data from these two experiments. An analysis of their data from various channels (individually, and combined) shows that the tension arises primarily from the $\nu_e$ appearance data, and is compounded by the $\bar{\nu}_\mu$ disappearance data. We provide an explanation for the tension based on parameter degeneracies. Apart from the analysis with the standard matter effect, we also analyse the data with the vacuum oscillation hypothesis. We find that vacuum oscillations fit the data as well as matter effects do; and also reduce the tension between the two experiments. We have also done a study of the future run of NO$\nu$A, T2K and DUNE in the context of establishing this tension with higher statistical significance.
\end{abstract}
\maketitle
\section{Introduction}

Neutrino oscillations have been well established by experimental data over the last three decades. Oscillation physics is now in the precision era, where the values of the various parameters that affect the oscillation probabilities are being measured to an unprecedented level of precision. Neutrino mixing is described by the unitary $3\times 3$ PMNS matrix:
\begin{equation}
U=\left[
\begin{array}{ccc}
c_{13}c_{12} & s_{12}c_{13} & s_{13}e^{-i\dcp}\\
-s_{12}c_{23}-c_{12}s_{23}s_{13}e^{i\dcp} & c_{12}c_{23}-s_{12}s_{23}s_{13}e^{i\dcp} & s_{23}c_{13}\\
s_{12}s_{23}-s_{13}c_{12}c_{23}e^{i\dcp} & -c_{12}s_{23}-s_{13}c_{23}s_{12}e^{i\dcp} & c_{23}c_{13}\\
\end{array}
\right],
\end{equation}
where $c_{ij}=\cos \theta_{ij}$ and $s_{ij}=\sin \theta_{ij}$. Neutrino oscillation probabilities depend on three mixing angles: $\tx$, $\ty$, and $\tz$; two independent mass squared differences $\ds=m_{2}^{2}-m_{1}^{2}$ and $\dl=m_{3}^{2}-m_{1}^{2}$, where $m_i$ is the mass of the neutrino mass eigenstate $\nu_i$; and a CP-violating phase, $\dcp$. Among these parameters, $\tx$ and $\ds$ have been measured in solar neutrino experiments \cite{Bahcall:2004ut, Ahmad:2002jz, Super-Kamiokande:1998qwk}, and in long baseline reactor neutrino experiment KamLAND \cite{KamLAND:2008dgz}. The long-baseline accelerator neutrino experiment MINOS \cite{Kyoto2012MINOS} measured $\sin^2 2\tz$ and $|\dl|$ by measuring $\nu_\mu$ survival probabilities\footnote{To be precise, MINOS used the $\nu_\mu$ disappearance channel to measure the effective atmospheric mixing parameters $\Delta_{\mu\mu}$ and $\theta_{\mu\mu}$~\cite{Nunokawa:2005nx,Raut:2012dm}}. $\ty$ has been measured in reactor neutrino experiments \cite{An:2012eh, Ahn:2012nd, Abe:2011fz}. The current unknowns are the sign of $\dl$, the octant of $\tz$ and the CP-violating phase $\dcp$. Depending on the sign of $\dl$, there can be two possible mass orderings of the neutrino mass eigenstates: normal hierarchy (NH) which implies $m_3>>m_2>m_1$, and inverted hierarchy (IH) which implies $m_2>m_1>>m_3$ \cite{deSalas:2018bym}. Similarly, for $\sin^2 2\tz$, there can be two possibilities: lower octant (LO), i.e. $\sin^2 \tz<0.5$ and higher octant (HO), i.e. $\sin^2 \tz>0.5$. 
The values of the mixing parameters are important as discriminators between new physics models that may be invoked to explain the tiny non-zero masses of the neutrinos~\cite{Albright:2006cw}. For models involving neutrinos of a Majorana nature, the mass ordering can have implications on the observability of processes like neutrinoless double beta-decay~\cite{Dolinski:2019nrj}. Further, the neutrino masses have significance for the energy density, and hence the expansion rate of the Universe. Constraints on (the sum of) neutrino masses also arise from cosmological observations~\cite{Hannestad:2010kz}, though these are model-dependent.

The currently running long-baseline accelerator neutrino oscillation experiments \nova \cite{Ayres:2004js} and T2K \cite{Itow:2001ee} are expected to measure these unknown quantities. In Neutrino 2020 conference, both the experiments released their latest data \cite{Himmel:2020, NOvA:2021nfi, Dunne:2020, T2K:2021xwb}. Each experiment has four data sets: 
\begin{itemize}
    \item [1.] $\nu_\mu$ disappearance data: $\nu_\mu \to \nu_\mu$, 
    \item [2.] $\bar{\nu}_\mu$ disappearance data: $\bar{\nu}_\mu \to \bar{\nu}_\mu$,
    \item [3.] $\nu_e$ appearance data: $\nu_\mu \to \nu_e$, and 
    \item [4.] $\bar{\nu}_e$ appearance data: $\bar{\nu}_\mu \to \bar{\nu}_e$.
\end{itemize}
In table~\ref{nova_t2k}, we have listed the best-fit values of the oscillation parameters measured by both the experiments. It is obvious that both the experiments agree on the best-fit value of $|\Delta_{32}|$ which essentially comes from the $\nu_\mu$ disappearance data. They also nearly agree on the best-fit value of $\sin^2 \tz$ at $1\,\sigma$. However there is a strong disagreement between the best-fit values of $\dcp$. Moreover, there is no overlap between the $1\, \sigma$ allowed regions on the $\sin^2 \tz-\dcp$ plane, as shown in Ref.~\cite{Himmel:2020}. However, the $3\, \sigma$ allowed regions are quite similar. Moreover, if IH is assumed to be the correct hierarchy, the allowed regions of the two experiments essentially coincide. Although both the experiments prefer NH over IH, the best-fit point for IH has a $\dchsq\simeq 1$ compared to the best-fit point at NH. In Ref.~\cite{Kelly:2020fkv}, it has been shown that the combined analysis of \nova and T2K has a minimum for IH instead of NH. The question of the physical origin of this discrepancy is still open. Joint analyses are also planned between the \nova and T2K collaborations with the aim of obtaining better constraints on oscillation parameters due to resolved degeneracy and to understand the non-trivial systematic correlations between them~\cite{Berns:2021iss}.

There have been studies attempting to explain this discrepancy with beyond standard model (BSM) physics. Ref.~\cite{Miranda:2019ynh} has tried to resolve the tension with non-unitary neutrino mixing while Ref.~\cite{Chatterjee:2020kkm, Denton:2020uda} have done the same with non-standard neutrino-matter NC interaction during propagation. In Ref.~\cite{Rahaman:2021leu} an effort has been made to resolve the tension with CPT violating Lorentz invariance violation. All these papers give a hint towards the presence of BSM physics in the present \nova and T2K data. The information obtained from these two experiments comes from the four oscillation channels listed above. If the tension between the data from \nova and T2K is indeed a signature of new physics, it is crucial to understand which channel or combination of channels is responsible for this difference. Alternately, if the difference is purely a statistical or systematic artifact, it is important to identify the channel that is the source of this effect. Thus it is important for both model-building and understanding the experimental systematics, to be certain about the origin of the tension. Previously a similar study was done with the 2018 data from \nova and T2K \cite{Nizam:2018got}.

In this paper, we try to identify the disagreement between the best-fit values of $\dcp$. Details of our analysis are discussed in section~\ref{analysis}. We present our result after analysing \nova and T2K data in section~\ref{result}. An explanation of the tension based on parameter degeneracy has been given in section~\ref{degeneracy}. In section~\ref{future}, future prospects of the long-baseline experiments are discussed and we have tried to determine if the tension can be established with greater statistical significance in future. The conclusions are presented in section~\ref{conclusion}.
\begin{table}
  \begin{center}
\begin{tabular}{|c|c|c|}
  \hline
  Parameters & NO$\nu$A&
  T2K\\
  
  \hline
  $\sin^2 \tz$ (NH) & $0.57^{+0.03}_{-0.04}$ & $0.512^{+0.045}_{-0.042}$\\
  \hline
  $\sin^2\tz$ (IH) & Not given & $0.500^{+0.050}_{-0.036}$\\
  \hline
  $\dcp$ (NH)  & $0.82\pi$ & $-2.14^{+0.90}_{-0.69}$ \\
  \hline
 $ \dcp/\pi$ (IH) & Not given & $-1.26^{+0.61}_{-0.69}$ \\
 \hline
  $ \frac{|\Delta_{32}|}{10^{-3}\, {\rm eV}^2}$ (NH)  & $2.41^{+0.07}_{-0.07}$ & $2.46^{+0.07}_{-0.07}$ \\
 \hline
 $ \frac{|\Delta_{32}|}{10^{-3}\, {\rm eV}^2}$ (IH)  & Not given & $2.50^{+0.18}_{-0.13}$ \\
 \hline
\end{tabular}
\end{center}
 \caption{Best-fit values of neutrino oscillation parameters as measured by \nova \cite{Himmel:2020, NOvA:2021nfi} and T2K \cite{Dunne:2020, T2K:2021xwb}}
  \label{nova_t2k}
\end{table}

\section{Analysis details}
\label{analysis}
\nova \cite{Ayres:2004js} consists of a 14 kt totally active scintillator detector (TASD), placed 810 km away from the Fermilab neutrino source (the NuMI beam), and is situated at $0.8^\circ$ off the beam axis. The neutrino flux peaks at $2$ GeV which is close to the oscillation maximum energy 1.4 GeV for NH and at 1.8 GeV for IH. \nova started taking data in 2014 and took data until the 2020 release \cite{Himmel:2020, NOvA:2021nfi} corresponding to $1.36 \times 10^{21}$ ($1.25 \times 10^{21}$) POT in $\nu$ ($\bar{\nu}$) mode.
The T2K experiment \cite{Itow:2001ee} consists of the $\nu_\mu$ beam from the J-PARC accelerator at Tokai and the water Cherenkov detector at Super-Kamiokande, which is 295 km away from the source and situated $2.5^\circ$ off-axis. The flux peaks at $0.7$ GeV, which is also close to the first oscillation maximum.  T2K started taking data in 2009, and until 2020 release of results these \cite{Dunne:2020, T2K:2021xwb} correspond to $1.97 \times 10^{21}$ ($1.63 \times 10^{21}$) POT in $\nu$ ($\bar{\nu}$) mode. 

To analyse the data, we kept $\sin^2\tx$ and $\ds$ at their best-fit values $0.304$ and $7.42\times 10^{-5}\, {\rm eV}^2$, respectively. We varied $\sin^2 2\ty$ in its $3\, \sigma$ range around its central value $0.084$ with $3.5\%$ uncertainty~\cite{Dohnal:2021rcr}. $\sin^2 \tz$ has been varied in its $3\, \sigma$ range $[0.41:0.62]$ (with $2\%$ uncertainty on $\sin^2 2\tz$ \cite{Esteban:2018azc}). We varied $|\Delta_{\mu \mu}|$ in its $3\, \sigma$ range around the MINOS best-fit value $2.32\times 10^{-3}\, {\rm eV}^2$ with $3\%$ uncertainty \cite{Nichol:2012}. $\Delta_{\mu \mu}$ is related with $\dl$ by the following relation \cite{Nunokawa:2005nx}
\begin{equation}
\Delta_{\mu \mu}= \sin^2 \tz \dl + \cos^2 \tx \Delta_{32}+\cos \dcp \sin 2\tx \sin \ty \tan \tx \ds.
\end{equation}
The CP-violating phase $\dcp$ has been varied in its complete range $[-180^\circ:180^\circ]$.

We calculated the theoretical event rates and the $\chi^2$ between data and theoretical event rates using GLoBES \cite{Huber:2004ka, Huber:2007ji}. The experimental data have been taken from \cite{Himmel:2020, NOvA:2021nfi, Dunne:2020, T2K:2021xwb}. To calculate the theoretical event rates, we fixed the signal and background efficiencies by matching with the Monte-Carlo simulations given by the collaborations \cite{Himmel:2020, NOvA:2021nfi, Dunne:2020, T2K:2021xwb}. Automatic bin-based energy smearing for generated theoretical events has been implemented within GLoBES~\cite{Huber:2004ka, Huber:2007ji} using a Gaussian smearing function
\begin{equation}
R^c (E,E^\prime)=\frac{1}{\sqrt{2\pi}}e^{-\frac{(E-E^\prime)^2}{2\sigma^2(E)}},
\end{equation}
where $E^\prime$ is the reconstructed energy. The energy resolution function is given by 
\begin{equation}
\sigma(E)=\alpha E+\beta \sqrt{E}+\gamma.
\end{equation}
For NO$\nu$A, we used $\alpha=0.11\, (0.09)$, $\beta=\gamma=0$ for electron (muon) like events \cite{NOvA:2018gge, NOvA:2019cyt}. For T2K, we used $\alpha=0$, $\beta=0.075$, $\gamma=0.05$ for both electron and muon like events. 

After the calculation of theoretical event rates, $\chi^2$s were calculated using the expression:
\begin{eqnarray}
%
\chi^2 &=& 2\sum_i \left\{
(1+z) N_i^{\rm th} - N_i^{\rm exp} + N_i^{\rm exp} 
\ln\left[ \frac{N_i^{\rm exp}}{(1+z) N_i^{\rm th}} \right]
\right\} + 2 \sum_j (1+z) N_j^{\rm th} + z^2 \,\, \nonumber \\
\label{poisionian}
\end{eqnarray}
where the index $i$ represents the bins for which $N_i^{\rm exp}\neq 0$, and $j$ represents the bins for which $N_j^{\rm exp} = 0$. The parameter $z$ defines the additional systematic uncertainties. For both NO$\nu$A and T2K, the relevant systematic uncertainties are
\begin{itemize}
    \item $5\%$ normalisation and $5\%$ energy callibration systematics uncertainty for $e$-like events, and
    \item $5\%$ normalisation and $0.01\%$ energy callibration systematics uncertainty for $\mu$-like events.
\end{itemize}

In calculating $\chi^2$ we added (the older, pre-2020) priors on $\sin^2 2\ty$, $\sin^2 2\tz$, and $|\Delta_{\mu \mu}|$, in cases where we have not included muon disappearance data. This is done to prevent `overcounting' of information obtained from the disappearance channels of the \nova and T2K dataset. In case of $\sin^2 2\ty$, and $\sin^2 \tz$, we used the priors from 2019 global analysis~\cite{nufit, Esteban:2018azc}, while for $|\Delta_{\mu \mu}|$ we have used the priors from the 2012 MINOS result \cite{Nichol:2012}. In all other cases, priors have been added on $\sin^2 2\ty$ only (to account for electron disappearance data from reactor neutrino experiments). After calculating $\chi^2$, we have marginalised it over all the oscillation parameters including the hierarchy, and subtracted it from the $\chi^2$s to calculate $\dchsq$.

While doing the analysis of future experiments, we calculated the true experimental event rates with \nova and T2K current best-fit points and calculated test event rates and $\dchsq$ as described above. In case of simulations, $\chi^2$ and $\dchsq$ are essentially equivalent.

\section{Analysis of the 2020 data}
\label{result}
Here we present the results after analysing \nova and T2K data. The results are presented in the $\sin^2\tz-\dcp$ plane. The value $\dchsq=0$ gives us the best-fit point in the parameter space, and it occurs for a particular hierarchy. In the figures, this point has been mentioned as "best fit". This is the global minimum point of the analysis. The minimum $\dchsq$ at the other hierarchy is non-zero, and it is mentioned as "NH/IH best fit" in the figures. The allowed regions are being shown according to the global minimum.

\subsection{Disappearance data}

\begin{figure}[h]
\centering
\includegraphics[width=1.0\textwidth]{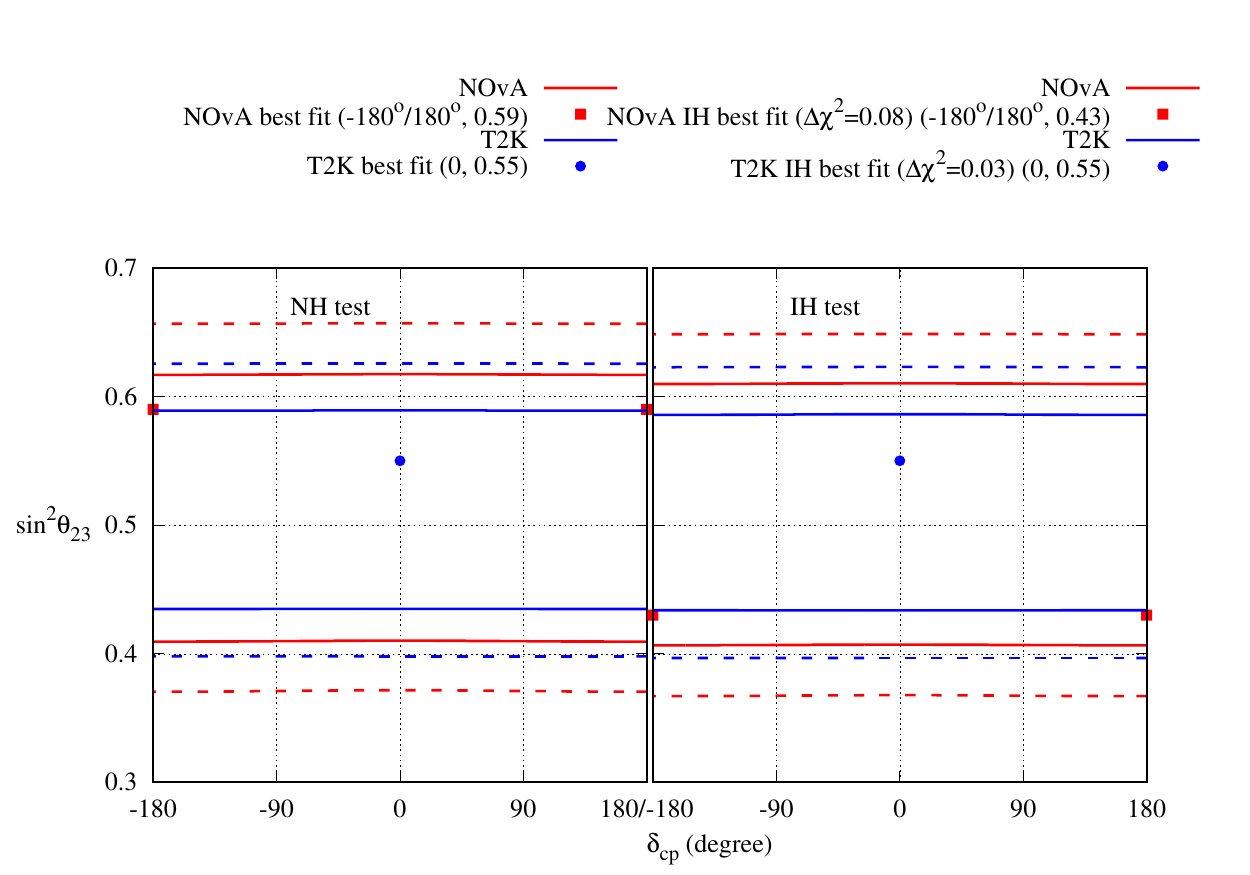}
\caption{\footnotesize{Allowed region in the $\sin^2 \tz-\dcp$ plane after analysing \nova and T2K $\nu_\mu$ disappearance
data.  The left (right) panel represents test hierarchy NH (IH). The red (blue) lines indicate the results for \nova (T2K). The solid (dashed) lines indicate the $1\, \sigma$ ($3\, \sigma$)
allowed regions. The minimum $\chi^2$ for \nova
(T2K) with 19 (35) bins is 16.12 (40.62) and it occurs at NH. The minimum $\dchsq$ for IH is mentioned in the figure itself.
}}
\label{nu-disapp}
\end{figure}

To begin with, we analysed the $\nu_\mu$ disappearance data of both the experiments. The minimum $\chi^2$ for \nova (T2K) with 19 (35) bins is 16.12 (40.62) and it occurs for NH. The result is presented in Fig.~\ref{nu-disapp}. We can see that the allowed regions for the two experiments are essentially same for the $\nu_\mu$ disappearance data.

\begin{figure}[h]
\centering
\includegraphics[width=1.0\textwidth]{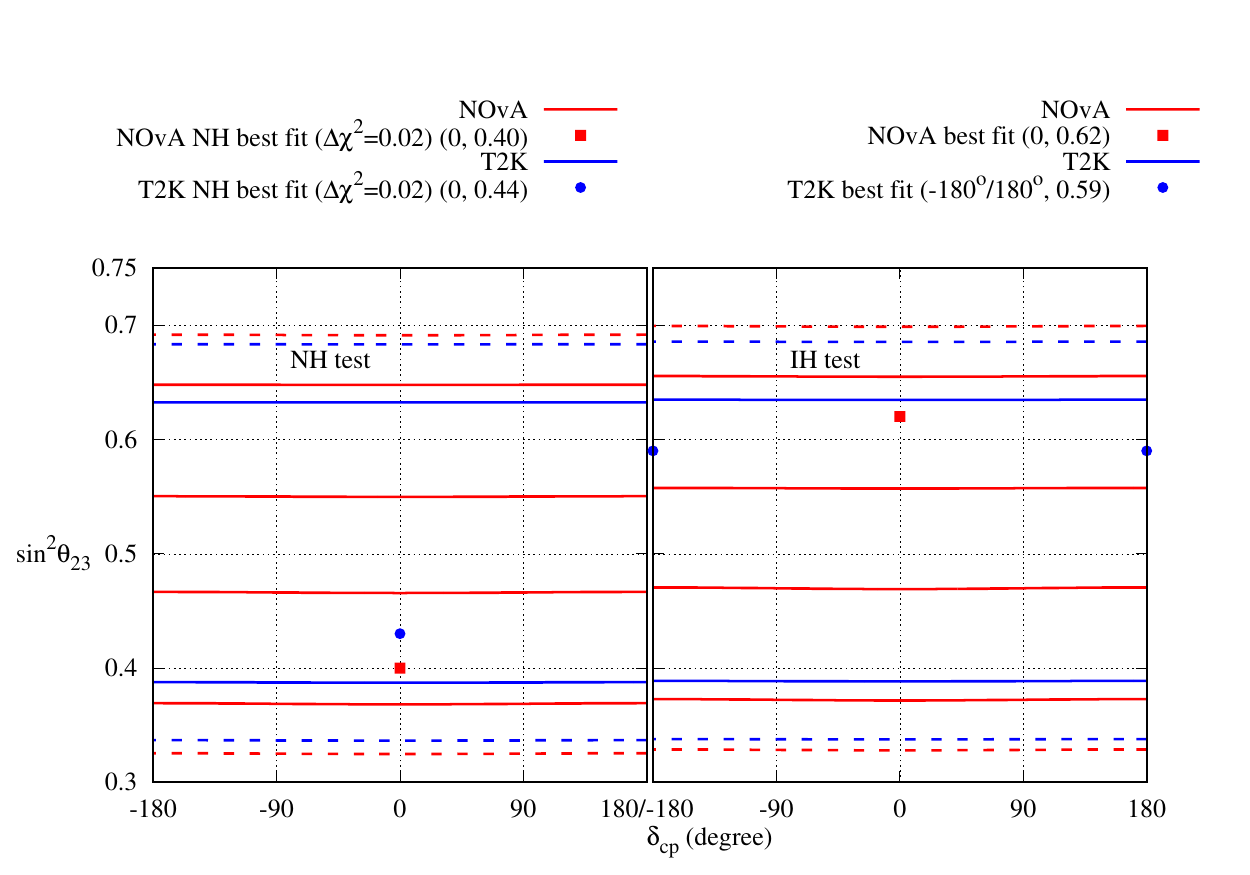}
\caption{\footnotesize{Allowed region in the $\sin^2 \tz-\dcp$ plane after analysing \nova and T2K $\bar{\nu}_\mu$ disappearance
data.  The left (right) panel represents test hierarchy NH (IH). The red (blue) lines indicate the results for \nova (T2K). The solid (dashed) lines indicate the $1\, \sigma$ ($3\, \sigma$)
allowed regions. The minimum $\chi^2$ for \nova
(T2K) with 19 (35) bins is 24.72 (34.64) and it occurs at IH. The minimum $\dchsq$ for NH is mentioned in the figure itself..
}}
\label{nubar-disapp}
\end{figure}

In Fig.~\ref{nubar-disapp}, we have represented the analysis of $\bar{\nu}_\mu$ disappearance data in the $\sin^2\tz-\dcp$ plane. The minimum $\chi^2$ for \nova
(T2K) with 19 (35) bins is 24.72 (34.64) and it occurs for IH. The minimum $\dchsq$ for NH is mentioned in the figure. We can see that T2K allows the whole region from LO to HO at the $1\, \sigma$ confidence level. However, \nova excludes regions around $\sin^2\tz=0.5$.

\begin{figure}[htbp]
\centering
\includegraphics[width=1.0\textwidth]{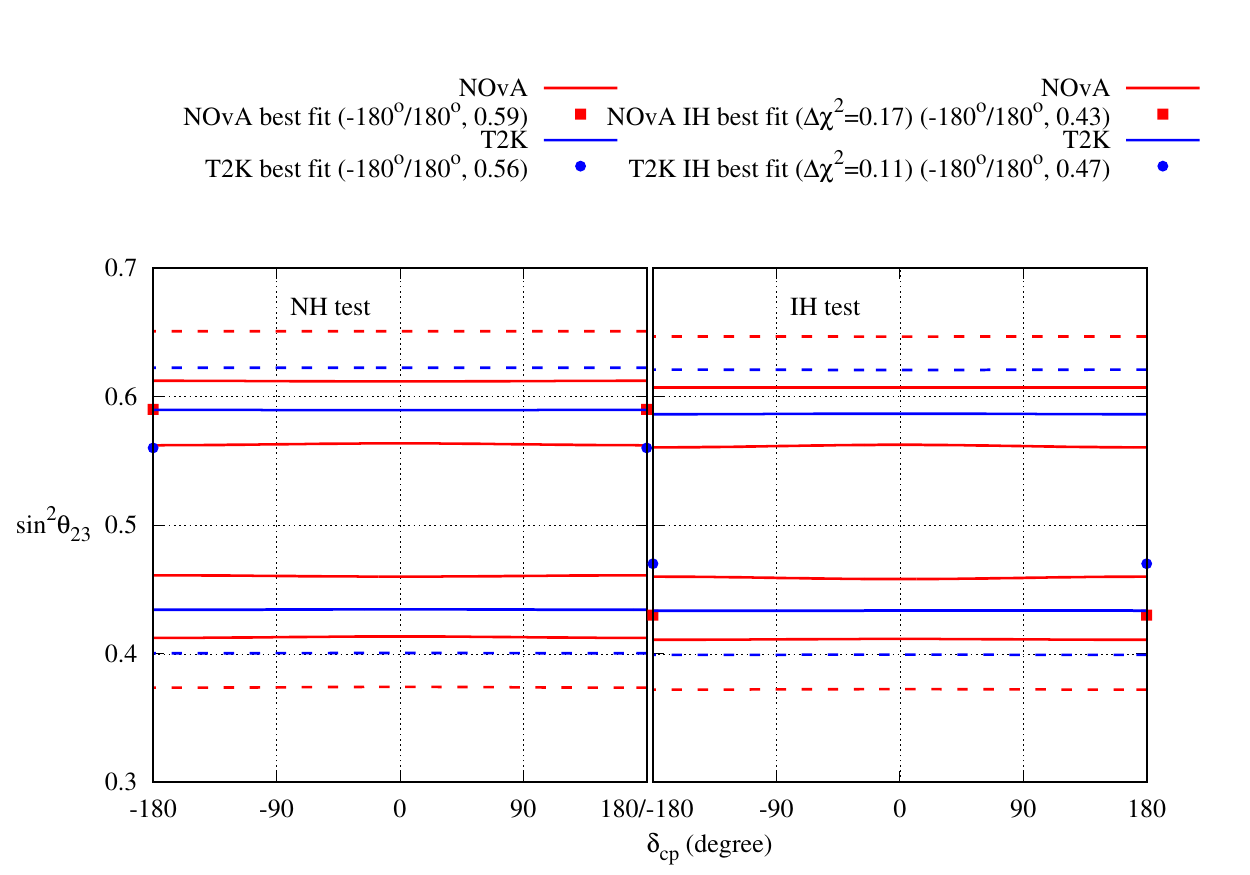}
\caption{\footnotesize{Allowed region in the $\sin^2 \tz-\dcp$ plane after analysing \nova and T2K $\nu_\mu$ disappearance and $\bar{\nu}_\mu$ disappearance
data. The left (right) panel represents test hierarchy NH (IH). The red (blue) lines indicate the results for \nova (T2K). The solid (dashed) lines indicate the $1\, \sigma$ ($3\, \sigma$)
allowed regions. The minimum $\chi^2$ for \nova
(T2K) with 38 (70) bins is 41.81 (75.86) and it occurs at NH. The minimum $\dchsq$ for IH is mentioned in the figure itself..
}}
\label{nu+nubar-disapp}
\end{figure}

For the combined analysis of $\nu_\mu+\bar{\nu}_\mu$ disappearance data, as shown in Fig.~\ref{nu+nubar-disapp}, the minimum $\chi^2$ for \nova
(T2K) with 38 (70) bins is 41.81 (75.86) and it occurs for NH. The minimum $\dchsq$ for IH is mentioned in the figure. It can be seen that both the experiments follow the same pattern seen in the $\bar{\nu}_\mu$ disappearance data. T2K allows the full range of $\sin^2\tz$ at $1\, \sigma$ C.L., while \nova continues to exclude regions around $\sin^2\tz=0.5$. From Figs.~\ref{nu-disapp}-\ref{nu+nubar-disapp}, it is obvious that there is no significant tension between the disappearance data of \nova and T2K.

\subsection{Appearance data}
\begin{figure}[htbp]
\centering
\includegraphics[width=1.0\textwidth]{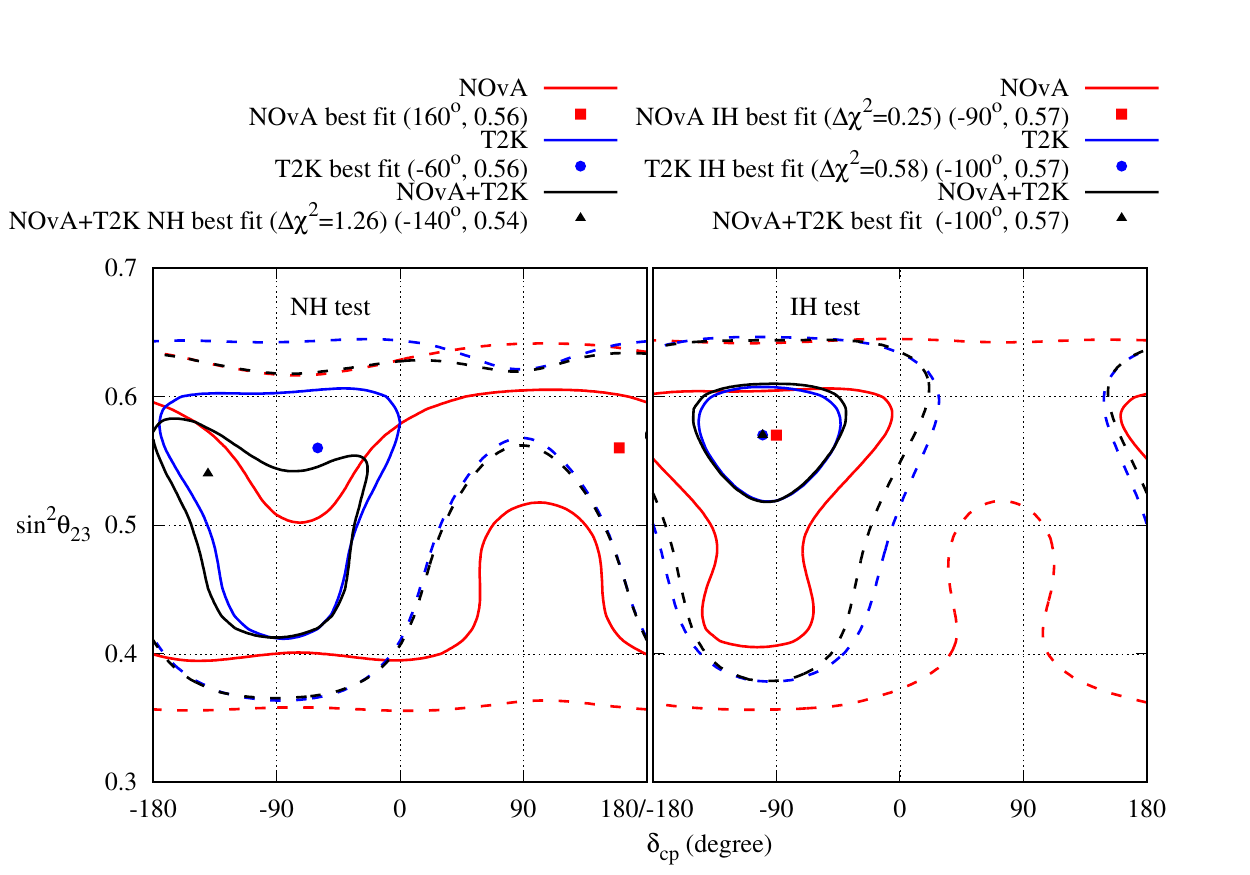}
\caption{\footnotesize{Allowed region in the $\sin^2 \tz-\dcp$ plane after analysing \nova and T2K $\nu_e$ appearance
data. The left (right) panel represents test hierarchy NH (IH). The red (blue) lines indicate the results for \nova (T2K). The solid (dashed) lines indicate the $1\, \sigma$ ($3\, \sigma$)
allowed regions. The minimum $\chi^2$ for \nova
(T2K) with 6 (11) bins is 3.69 (12.24) and it occurs at NH. The minimum $\dchsq$ for IH is mentioned in the figure itself. The minimum $\chi^2$ for the combined analysis is 16.76 for 17 energy bins and it occurs at IH. The minimum $\dchsq$ for NH is mentioned in the figure.
}}
\label{nu-app}
\end{figure}

In the next step, we analysed the $\nu_e$ appearance data from each of the two experiments. We have also done the combined analysis of the $\nu_e$ appearance data from both of the experiments. The minimum $\chi^2$ for \nova (T2K) with 6 (11) bins is 3.69 (12.24) and it occurs at NH. The minimum $\dchsq$ for IH is mentioned in the Fig.~\ref{nu-app}. For the combined analysis. the minimum $\chi^2$ is 16.76 for 17 energy bins and it occurs at IH.
The result has been shown in Fig.~\ref{nu-app}. From the left panel (test hierarchy NH), we can see that about $50\%$ of the $1\, \sigma$ allowed region of T2K is allowed by the $1\, \sigma$ allowed region of NO$\nu$A, but only a small portion of the $1\, \sigma$ allowed region of \nova is allowed by T2K $1\, \sigma$ allowed region. The two experiments have the same allowed region at $3\, \sigma$, when NH is the test hierarchy. When IH is the test hierarchy, which is nearly degenerate with NH, the best-fit points of the two experiments coincide.
 
\begin{figure}[H]
\centering
\includegraphics[width=1.0\textwidth]{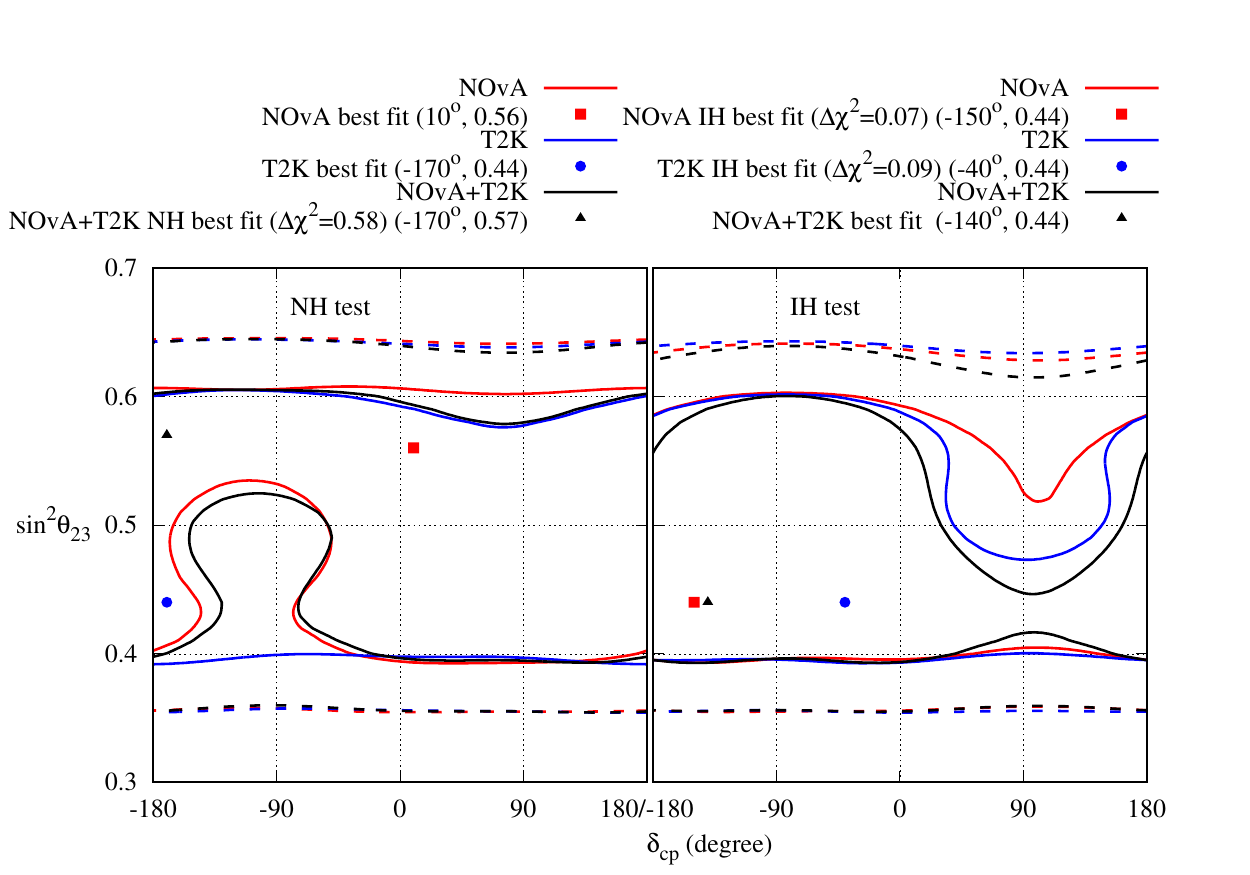}
\caption{\footnotesize{Allowed region in the $\sin^2 \tz-\dcp$ plane after analysing \nova and T2K $\bar{\nu}_e$ appearance
data.  The left (right) panel represents test hierarchy NH (IH). The red (blue) lines indicate the results for \nova (T2K). The solid (dashed) lines indicate the $1\, \sigma$ ($3\, \sigma$)
allowed regions. The minimum $\chi^2$ for \nova
(T2K) with 6 (7) bins is 2.91 (7.50) and it occurs at NH. The minimum $\dchsq$ for IH is mentioned in the figure itself. The minimum $\chi^2$ for the 13 energy bins of the combined analysis is 10.83 and it occurs at IH.
}}
\label{nubar-app}
\end{figure}

 For $\bar{\nu}_e$ appearance, the minimum $\chi^2$ for \nova
(T2K) with 6 (7) bins is 2.91 (7.50) and it occurs at NH. The minimum $\dchsq$ for IH is mentioned in the Fig.~\ref{nubar-app}. For the combined analysis of both the experiments, the minimum $\chi^2$ for 13 energy bins is 10.83 and it occurs at IH. The experiments have very different best-fit points for both NH and IH, but the $1\, \sigma$ and $3\, \sigma$ allowed regions coincide.

\begin{figure}[htbp]
\centering
\includegraphics[width=1.0\textwidth]{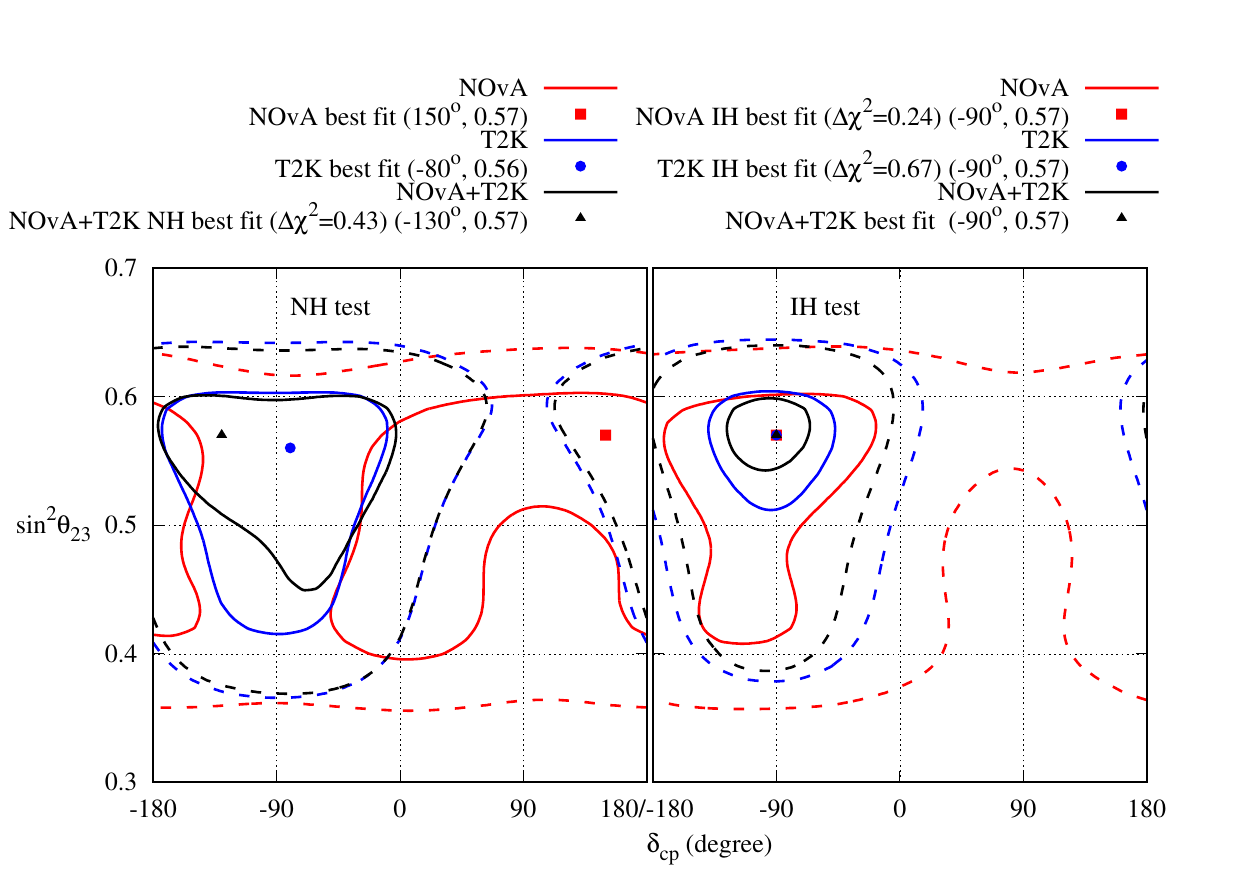}
\caption{\footnotesize{Allowed region in the $\sin^2 \tz-\dcp$ plane after analysing \nova and T2K $\nu_e$ appearance and $\bar{\nu}_e$ appearance
data. The left (right) panel represents test hierarchy NH (IH). The red (blue) lines indicate the results for \nova (T2K) and the black line indicates the allowed region for the combined analysis. The solid (dashed) lines indicate the $1\, \sigma$ ($3\, \sigma$)
allowed regions. The minimum $\chi^2$ for \nova
(T2K) with 12 (18) bins is 6.77 (20.23) and it occurs at NH. The minimum $\dchsq$ for IH is mentioned in the figure itself. The minimum $\chi^2$ for combined analysis is 27.89 for 30 bins and it occurs at IH. The minimum $\dchsq$ for NH is mentioned in the figure itself.
}}
\label{nu+nubar-app}
\end{figure}

In the next step, we have analysed the combined $\nu_e$ and $\bar{\nu}_e$ appearance data from each of the two experiments. The minimum $\chi^2$ for \nova
(T2K) with 12 (18) bins is 6.77 (20.23) and it occurs at NH. The minimum $\dchsq$ for IH is mentioned in the Fig.~\ref{nu+nubar-app}. If we combine the data from \nova and T2K, the best-fit point occurs at IH. The minimum $\chi^2$ for 30 bins is 27.89. From Fig.~\ref{nu+nubar-app}, it is evident that \nova and T2K $1\, \sigma$ regions are mutually exclusive for test NH but they have complete overlap for test IH. 

Taking a closer look here, for NH, the $\nu_e$ appearance data of T2K rule out $\dcp$ in upper half plane (UHP, $0<\dcp<180^\circ$) and allow only the lower half plane (LHP, $-180^\circ<\dcp<0$) of $\dcp$. On the other hand, the $\bar{\nu}_e$ appearance data of T2K do not care whether $\dcp$ is in UHP
or LHP and allow $\tz$ in both LO and HO. Hence, the combined effect of $\nu_e$ appearance and $\bar{\nu}_e$ appearance limit T2K $1\, \sigma$ region to be in LHP. For the same hierarchy, the $\nu_e$ appearance data of \nova allow UHP of $\dcp$. In the LHP, \nova $\nu_e$ appearance data show a preference for $\tz$ in LO over HO. For the $\bar{\nu}_e$ appearance, \nova data favour $\dcp$ in LHP when $\tz$ is in HO rather than LO. Therefore, \nova $\nu_e$ and $\bar{\nu}_e$ appearance data are mutually exclusive to each other when $\dcp$ is in LHP. Because of this the combined $\nu_e$ and $\bar{\nu}_e$ appearance data of \nova want the $\dcp$ to be in UHP and allow only a very small range of $\dcp$ in LHP. Thus, there is only a very tiny overlap of $1\, \sigma$ allowed regions of the two experiments. 
\subsection{Complete data}
\begin{figure}[htbp]
\centering
\includegraphics[width=1.0\textwidth]{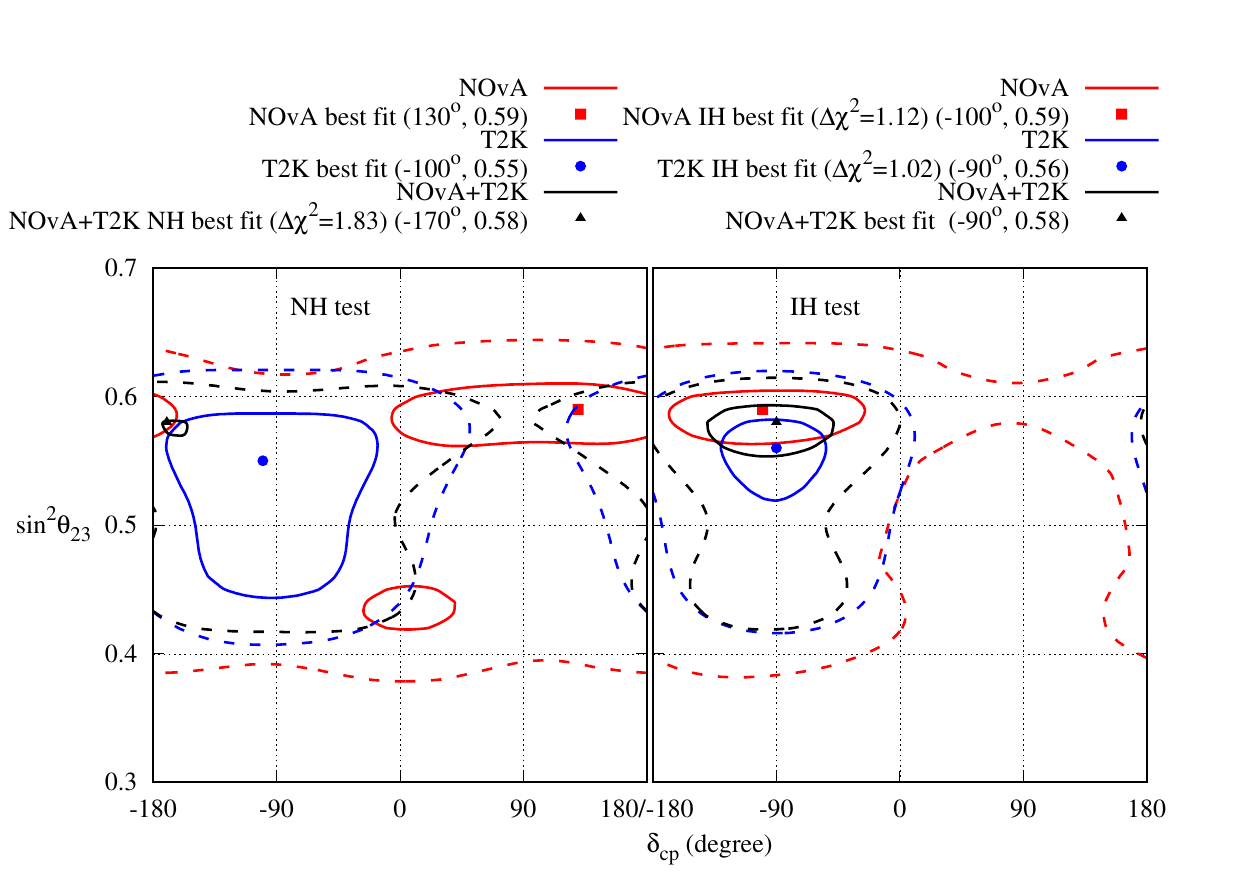}
\caption{\footnotesize{Allowed region in the $\sin^2 \tz-\dcp$ plane after analysing \nova and T2K complete data set. The left (right) panel represents test hierarchy NH (IH). The red (blue) lines indicate the results for \nova (T2K)
and the black line indicates the combined analysis of both. The solid (dashed) lines indicate the $1\, \sigma$ ($3\, \sigma$)
allowed regions. The minimum $\chi^2$ for \nova
(T2K) with 50 (88) bins is 48.65 (95.85) and it occurs at NH. For the combined analysis, the minimum $\chi^2$ with 138 bins is 147.14.
}}
\label{nu+nubar-app+disapp}
\end{figure}

Finally we have done a combined analysis of all four data pieces from each of the two experiments. The minimum $\chi^2$ for \nova (T2K) with 50 (88) bins is 48.65 (95.85) and it occurs at NH. For the combined analysis of the two experiments,
minimum $\chi^2$ for 138 bins is 147.14 and it occurs at IH. From Fig.~\ref{nu+nubar-app+disapp}, it is obvious that the NH $1\, \sigma$ region of T2K is more or less similar to that of $\nu_e+\bar{\nu}_e$ appearance data. But the NH $1\, \sigma$ region of \nova shrinks substantially compared to that of the appearance data only. This is because of the fact that the disappearance data of \nova rule out $\sin^2\tz \simeq 0.5$. Thus the inclusion of disappearance data make the disagreement between the two experiments stronger. For IH, there is a strong overlap at $1\, \sigma$. At $3\, \sigma$, there is a strong overlap for both NH and IH. 

\subsection{Analysing the data with vacuum oscillation hypothesis}
 The matter effect is included in the neutrino oscillation probabilities through the Wolfenstein matter term \cite{msw1}

\begin{equation}
    A=2\sqrt{2}G_FN_eE=0.76\times 10^{-4} \rho ({\rm g/cc})E(GeV)\, \, ({\rm in\, \, eV^2}),
\end{equation}
where $E$ is the neutrino beam energy, $N_e$ is the electron density in matter, and $\rho$ is the earth matter density. Matter effect plays a crucial role in solving the solar neutrino problem \cite{Mikheyev:1985zog, Mikheev:1986wj, Bahcall:2004ut}. Existence of matter effect has been established at more than $5\, \sigma$ C.L. for oscillations driven by $\ds$ \cite{Fogli:2005cq}. However, there is still no significant evidence of matter effect in the case of oscillations driven by $\dl$ \cite{Bharti:2020gnu}. It is expected that in case of atmospheric neutrinos, the survival probabilities $\pmumu$ and $\pmumubar$ will undergo significant changes due to matter effect \cite{Super-Kamiokande:2017yvm}. However, at present Super-Kamiokande can only distinguish between vacuum oscillation and matter effect at $2\, \sigma$ \cite{Super-Kamiokande:2017yvm}. For baselines of the order of 1000 km, changes in survival probabilities due to matter effect are negligible \cite{Gandhi:2007td}. However, the oscillation probabilities $\pmue$ and $\pmuebar$ undergo large changes due to matter effect. In cases of experiments like \nova and T2K, the sensitivities to mass hierarchy, $\dcp$ and octant of $\tz$ come from measuring these two oscillation probabilities \cite{Lipari:1999wy, Narayan:1999ck}. Since T2K has a smaller baseline (and correspondingly, lower energy and matter density), the change in oscillation probabilities due to matter effect is also smaller. Therefore, the measured values of the unknown oscillation parameters in T2K do not depend significantly on whether matter effect is considered or not \cite{Bharti:2020gnu}. However, for \nova, the changes in oscillation probabilities due to matter effect are comparable to the changes due to the shift of $\dcp$ value by $90^\circ$ \cite{Bharti:2018eyj}. That is why the $\dcp$ values measured in \nova are highly dependent on whether matter effect is considered or not. Just like the oscillations driven by two different mass-squared differences have been established, it is extremely important to establish the matter effects in the case of oscillations driven by $\dl$ at the same level of significance as those driven by $\ds$.

\begin{figure}[htbp]
\centering
\includegraphics[width=1.0\textwidth]{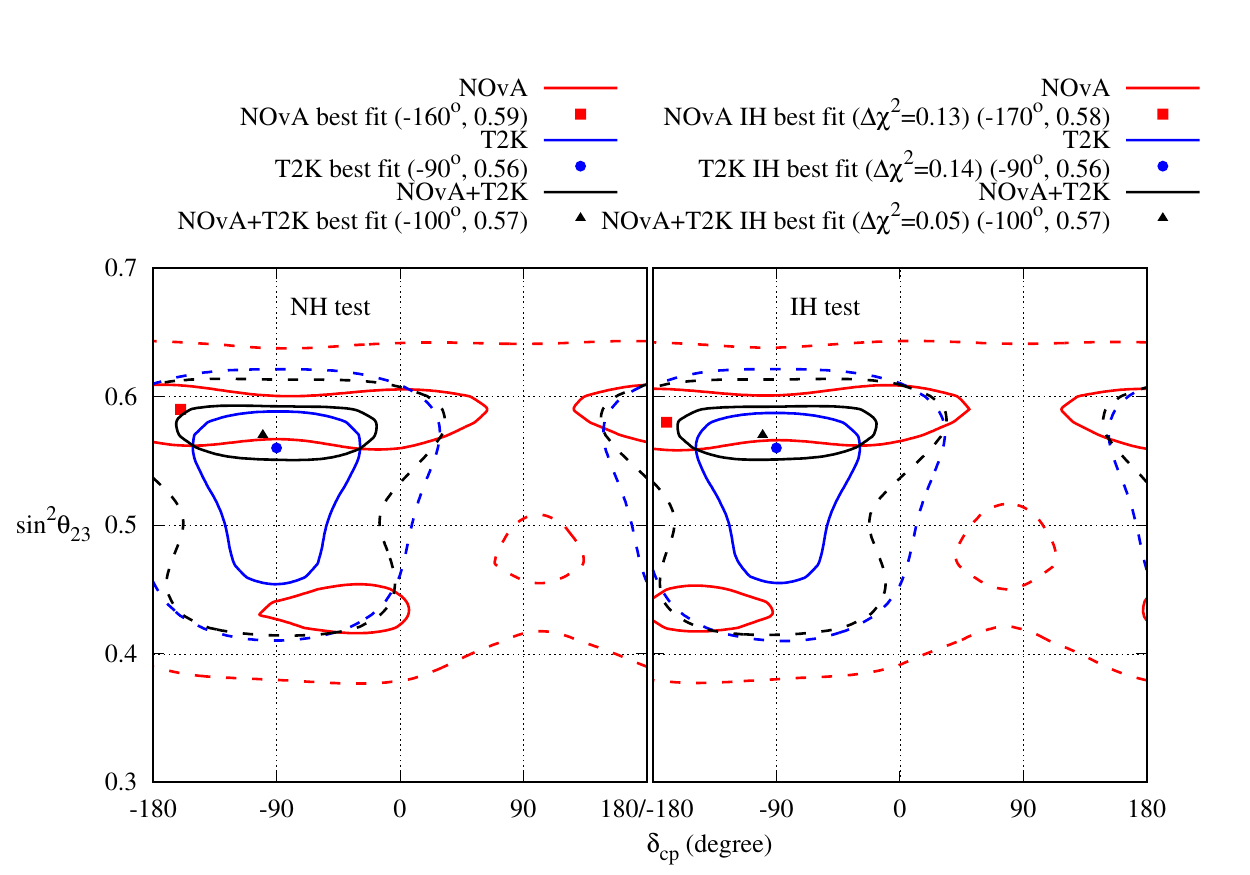}
\caption{\footnotesize{Allowed region in the $\sin^2 \tz-\dcp$ plane after analysing \nova and T2K complete data set with vacuum oscillation hypothesis. The left (right) panel represents test hierarchy NH (IH). The red (blue) lines indicate the results for \nova (T2K)
and the black line indicates the combined analysis of both. The solid (dashed) lines indicate the $1\, \sigma$ ($3\, \sigma$)
allowed regions. The minimum $\chi^2$ for \nova
(T2K) with 50 (88) bins is 48.84 (95.85). For the combined analysis, the minimum $\chi^2$ with 138 bins is 147.02.  
}}
\label{vacuum}
\end{figure}

In Ref.~\cite{Bharti:2020gnu}, it was shown that the 2019 data of \nova and T2K could not disfavour vacuum oscillation even at $1\, \sigma$.
We have done an analysis of the full 2020 data from each of the two experiments with vacuum oscillation hypothesis and the results are shown in Fig.~\ref{vacuum}. As expected, if the data are analysed with vacuum oscillation hypothesis, a distinction between NH and IH cannot be made. The minimum $\chi^2$ for \nova (T2K) with 50 (88) bins is 48.94 (95.85). The minimum $\chi^2$ for the combined analysis with 138 bins is 147.02. Therefore, vacuum oscillation gives as good fit to the data as oscillation with matter effect. From Fig.~\ref{vacuum}, it can be observed that if we use vacuum oscillation hypothesis, there is a compatibility between the two experiments at $1\, \sigma$.

\subsection{Summary of the analysis}
\begin{table}
  \begin{center}
\begin{tabular}{|c|c|c|c|}
  \hline
  Mode & NO$\nu$A best-fit&
  T2K best-fit & NO$\nu$A+T2K best-fit\\
  \hline
  $\nu_\mu$ disappearance & NH-$0.59$-$-180^\circ/180^\circ$  & NH-$0.55$-$0$&\\
  \hline
  $\bar{\nu}_\mu$ disappearance & IH-$0.62$-$0$ & IH-$0.59$-$-180^\circ/180^\circ$&\\
  \hline
  $\nu_\mu+\bar{\nu}_\mu$ disappearance  & NH-$0.59$-$-180^\circ/180^\circ$ & NH-$0.56$-$-180^\circ/180^\circ$& \\
  \hline
 $\nu_e$ appearance & NH-$0.56$-$160^\circ$ & NH-$0.56$-$-60^\circ$& IH-$0.57$-$-100^\circ$ \\
 \hline
  $\bar{\nu}_e$ appearance  & NH-$0.56$-$10^\circ$ & NH-$0.44$-$-170^\circ$&IH-$0.44$-$-140^\circ$ \\
 \hline
 $ \nu_e+\bar{\nu}_e$ appearance  & NH-$0.57$-$150^\circ$ & NH-$0.56$-$-80^\circ$& IH-$0.57$-$-90^\circ$ \\
 \hline
Complete data (matter effect)  & NH-$0.59$-$130^\circ$ & NH-$0.55$-$-100^\circ$ & IH-$0.58$-$-90^\circ$ \\
 \hline
Complete data (vacuum oscillations)  & $0.59$-$-160^\circ$ & $0.56$-$-90^\circ$& $0.57$-$-100^\circ$ \\
 \hline
\end{tabular}
\end{center}
 \caption{Best-fit values of \nova and T2K for different modes. The best-fit values have been presented as hierarchy-$\sin^2\tz$-$\dcp$. For vacuum oscillation hypothesis, we did not mention the hierarchy, as vacuum oscillation cannot distinguish between the two hierarchies.}
  \label{bestfit}
\end{table}
In table \ref{bestfit}, we have noted down the best-fit values of \nova and T2K for different modes. It is evident that there is no significant tension between the two experiments in the $\nu_\mu$ and $\bar{\nu}_\mu$ disappearance data \footnote{Note that, disappearance probabilities are insensitive to the value of $\dcp$. Therefore, we disregard the best-fit values of this parameter in the first three rows.}. However, the \nova $\bar{\nu}_\mu$ disappearance data rule out $\sin^2\tz$ values around $0.5$, as seen from figs.~\ref{nubar-disapp} and \ref{nu+nubar-disapp}. Although the best-fit values of \nova and T2K are far apart from each other for the $\bar{\nu}_e$ appearance data, there is no significant tension between the allowed parameter space of the two experiments. The $\bar{\nu}_e$ data for both the experiments are inconclusive about $\dcp$ and $\sin^2\tz$ values and the allowed regions of both the experiments overlap with each other, as seen in Fig.~\ref{nubar-app}. The tension arises mostly from $\nu_e$ appearance data, as can be seen in Fig.~\ref{nu-app}. The best-fit points of the two experiments are far apart and there is only a small overlap between their allowed regions at $1\, \sigma$ C.L. When the complete data set of the two experiments are analysed, the tension increases. The best-fit points are close to the ones for $\nu_e$ appearance data. We also saw that vacuum oscillation gives as good fit to the data as matter effect and the tension is reduced when the data are analysed with vacuum oscillation hypothesis. 

It should be noted that the sensitivity to the unknown parameters mostly come from the $\nu_e$, and $\bar{\nu}_e$ appearance data. The tension between the two experiments mostly arises from the $\nu_e$ appearance channel. In the next section we give the reason behind the tension between \nova and T2K in terms of effects of oscillation parameters on oscillation probabilities.
\section{Explaining the tension with parameter degeneracy}
\label{degeneracy}
We will try to explain the tension between the two experiments in terms of the effects of oscillation parameters degeneracy on the (anti-) electron neutrino appearance events. To do so, let us first look at the $\nu_\mu \to \nu_e$ oscillation probability for a neutrino of energy $E$ traversing a distance $L$ \cite{Cervera:2000kp}: 
\begin{eqnarray}
  &\pme& \simeq \sin^2 2 \ty \sin^2 \tz\frac{\sin^2\dhat(1-\ahat)}{(1-\ahat)^2}\nonumber\\
  &+&\alpha \cos \ty \sin2\tx \sin 2\ty \sin 2\tz \cos(\dhat+\dcp)\nonumber\\
 &&\frac{\sin\dhat \ahat}{\ahat}
  \frac{\sin \dhat(1-\ahat)}{1-\ahat},
  \label{pme}
   \end{eqnarray}
    where $\alpha=\frac{\ds}{\dl}$, $\dhat=\frac{\dl L}{4E}$
and $\ahat=\frac{A}{\dl}$. The antineutrino oscillation probability $\pmebar$ can be obtained by changing the sign of $A$ and $\dcp$ in eq.~\ref{pme}. The oscillation probability formula shows the dependence on mass hierarchy, octant of $\tz$ and $\dcp$.

From eq.~\ref{pme}, it can be easily seen that the dominant term in $\pme$ is proportional to $\sin^2 2\ty$, and thus this probability is rather small. Matter effect can enhance (suppress) the probability if $\dl$ is positive (negative). The situation is reversed for $\pmebar$. The dominant term is also proportional to $\sin^2\tz$. When $\tz$ is in LO (HO), i.e. $\sin^2 \tz<0.5$ ($\sin^2\tz>0.5$), both $\pme$ and $\pmebar$ is suppressed (enhanced) relative to the maximal mising, i.e. $\sin^2 \tz=0.5$. $\dcp$ sensitivity of the experiments comes from the second term which is proportional to $\alpha\approx 0.03$. When $\dcp$ is in LHP, $\pme$ ($\pmebar$) is larger (smaller), compared to the CP conserving case. However, when $\dcp$ is in UHP, $\pme$ ($\pmebar$) is smaller (larger), compared to the CP conserving case. Since each of the unknowns can choose two different values, there are eight possible combinations of the unknowns. Any given value of $\pme$ can be reproduced with any combination of the unknowns by choosing the value of $\ty$ appropriately. Thus there is an eight-fold degeneracy \cite{Fogli:1996pv, BurguetCastell:2001ez, Minakata:2001qm, Mena:2004sa, Prakash:2012az, Meloni:2008bd, Agarwalla:2013ju, Nath:2015kjg, Bora:2016tmb, Barger:2001yr, Kajita:2006bt} in the expression of $\pme$ and $\pmebar$ if $\sin^2 2\ty$ is not known precisely. The recent precision measurement of $\ty$ breaks this eight-fold degeneracy into $(1+3+3+1)$ pattern \cite{Bharti:2018eyj}. 

To understand the tension between \nova and T2K, we will first calculate the expected $\nu_e$ and $\bar{\nu}_e$ appearance event rates (signal+background) of \nova and T2K for vacuum oscillation, $\sin^2 \tz=0.5$ and $\dcp=0$. We label this combination as $000$. This is our benchmark point. Parameter values which increase $\pme$ will be labeled as '$+$' and those which decrease $\pme$ will be labeled as '$-$'. Therefore,
\begin{itemize}
    \item [1.] NH will be labelled as $+$, and IH will be labelled as $-$.
    \item [2.] $\tz$ in HO will be labelled as $+$, and $\tz$ in LO will be labelled as $-$
    \item [3.] $\dcp$ in LHP will be labelled as $+$, and $\dcp$ in UHP will be labelled as $-$.
\end{itemize}
Although $\pmebar$ is decreased (increased) by NH (IH), $\dcp$ in LHP (UHP), we will continue to label them as $+$ ($-$) for anti-neutrino as well. 

After calculating the expected event rates with the benchmark parameter values, we calculate the expected $\nu_e$ and $\bar{\nu}_e$ event rates with one unknown parameter changed at a time and then all three parameters changed together in the same direction.In tables \ref{nova_events} and \ref{t2k_events}, we have listed the expected $\nu_e$ and $\bar{\nu}_e$ event numbers of \nova and T2K respectively for different combinations of unknown oscillation parameters. The observed $\nu_e$ ($\bar{\nu}_e$) event numbers for \nova and T2K are respectively $82$ ($33$) and $107$ ($15$).

From table~\ref{nova_events}, we see that at the benchmark point $000$ the expected $\nu_e$ ($\bar{\nu}_e$) appearance event rate for \nova is 76.14 (32.93). The observed $\nu_e$ ($\bar{\nu}_e$) event rate is 82 (33). Therefore, there is a moderate excess in the $\nu_e$ appearance channel. This excess can be explained if change due to two unknown parameters cancel each other and there is a boost from the third unknown parameter. In terms of the labels defined above, the possible solutions are: (i) $++-$, (ii) $+-+$, and (iii) $-++$. It is to be noted that $000$ and $+00$ labels also lie in the $1\, \sigma$ range of the observed electron events. As for $\bar{\nu}_e$ appearance channel, the observed number of events is consistent with the expected number of events at $000$. However, due to the lack of statistics in the $\bar{\nu}$ data, all other possible combinations are also allowed, except $+-+$, and $-+-$ (these two combinations lead to the minimum and maximum number of expected event rates in the $\bar{\nu}_e$ appearance channel respectively). Therefore the analysis of the \nova data in entirety gives solution in the form of $++-$ and $-++$. These features are visible in figures \ref{nu-app}-\ref{nu+nubar-app}. $\bar{\nu}_\mu$ disappearance data of \nova rule out values around $\sin^2\tz=0.5$ and hence, when the appearance data are combined with the disappearance data of \nova, the $1\, \sigma$ allowed regions are closer to $++-$, and $-++$ labels. Apart from that there is a small allowed region around $+-0$ and $++0$.

\begin{figure}[htbp]
\centering
\includegraphics[width=1.0\textwidth]{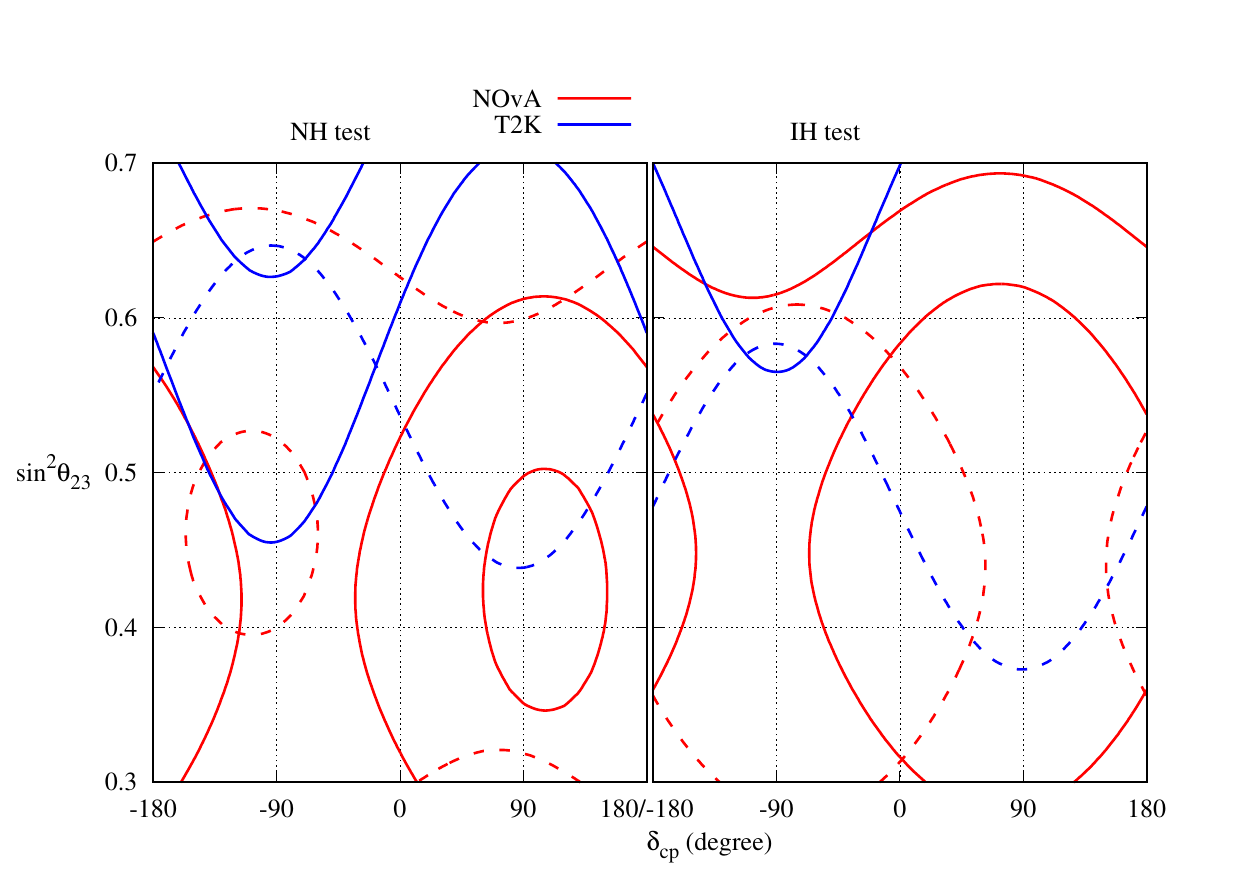}
\caption{\footnotesize{Contour plot on the $\sin^2\tz-\dcp$ showing the parameter values for which the expected event numbers fall within $N_{\rm obs}\pm \sqrt{N_{\rm obs}}$ range. $N_{\rm obs}$ for \nova (T2K) $\nu_e$ and $\bar{\nu}_e$ are 82 (107) and 33 (15) respectively. In the figure, the red (blue) lines indicate \nova (T2K) and the solid (dashed) lines indicate $\nu_e$ ($\bar{\nu}_e$) expected events in the mentioned range.
}}
\label{event-cont}
\end{figure}

For T2K, the expected $\nu_e$ ($\bar{\nu}_e$) appearance event rate at the benchmark point is 78 (19) (see table~\ref{t2k_events}). Choosing NH leads to a moderate $7-8\%$ boost for T2K $\nu_e$ appearance events. $\nu_e$ appearance events get a large boost if $\tz$ is high in higher octant. However, the disappearance data do not allow $\sin^2\tz>0.59$. Given that only around $20\%$ boost is possible due to hierarchy and octant, T2K $\nu_e$ appearance data firmly anchors around $\dcp\approx -90^\circ$. For the choice of NH, $\sin^2\tz=0.59$ and $\dcp=-90^\circ$, the expected $\nu_e$ appearance events for T2K is $113$, which is within $1\,\sigma$ range of the observed $\nu_e$ event rates $107$.  At best-fit point given in table~\ref{nova_t2k}, the expected number of events is $103$. Therefore, the observation of $107$ electron events seems to preclude the consideration of $\dcp$ values other than $\dcp=-90^\circ$. All other data: T2K $\bar{\nu}_e$ appearance events, \nova $\nu_e$ and $\bar{\nu}_e$ appearance events allow other possible $\dcp$ values, but T2K $\nu_e$ data is anchored to $\dcp\approx -90^\circ$. This is one of the biggest reasons for the large difference between the best-fit points of \nova and T2K and the non-availability of overlap between the $1\, \sigma$ allowed regions of the two experiments. As discussed above, the present $\nu_e$ appearance events of \nova can be obtained with the combinations NH-$\tz$ in HO-$\dcp$ in UHP ($++-$) or IH-$\tz$ in HO-$\dcp$ in LHP ($-++$) . But the T2K $\nu_e$ appearance data can only be obtained with $\tz$ in HO and $\dcp$ values around $-90^\circ$ for both NH and IH ($+++$ and $-++$). These features can be easily seen from Fig.~\ref{event-cont}. In this figure, we have shown the contour on $\sin^2\tz-\dcp$ plane, for which the expected $\nu_e$ and $\bar{\nu}_e$ event numbers fall within the range $N_{\rm obs}\pm \sqrt{N_{\rm obs}}$ for both \nova and T2K. Here $N_{\rm obs}$ is the observed event numbers in the experiments. It can be seen that for T2K, the expected $\nu_e$ event numbers are compatible with $N_{\rm obs}\pm \sqrt{N_{\rm obs}}$ when $\dcp$ is in LHP and $\tz$ mostly in HO, when NH is the true hierarchy and when $\sin^2\tz$ is in its global $3\, \sigma$ range $[0.41:0.63]$ \cite{nufit, Esteban:2018azc}. For IH also, expected $\nu_e$ event numbers for T2K fall within this range for $\dcp\approx-90^\circ$ and $\tz$ in HO. For NO$\nu$A, however, the expected $\nu_e$ event numbers can fall in the mentioned range for $\dcp$ in UHP for NH and around $-90^\circ$ for IH. This role played by the $\nu_e$ event numbers in the tension between \nova and T2K is also visible in Fig.~1 of Ref.~\cite{Esteban:2020cvm} and Fig.~8 of Ref.~\cite{Capozzi:2021fjo}.

\begin{table}
  \begin{center}
\begin{tabular}{|c|c|c|c|}
  \hline
  Hierarchy-$\sin^2\tz$-$\dcp$ & Label&
  $\nu_e$ Appearance events&$\bar{\nu}_e$ Appearance events\\
  
  \hline
   Vacuum-$0.5$-$0$ & $000$ & $76.14$ & $32.93$ \\
  \hline
  NH-$0.5$-$0$ & $+00$ & $86.16$& $29.24$\\
  \hline
  NH-$0.59$-$0$  & $++0$ & $99.48$ & $33.67$ \\
  \hline
 NH-$0.59$-$-90^\circ$ & $+++$ & $109.24$ & $29.06$ \\
 \hline
  NH-$0.59$-$+90^\circ$ & $++-$ & $84.47$ & $36.32$ \\
 \hline
 NH-$0.41$-$-90^\circ$  & $+-+$ & $91.98$ & $24.91$ \\
 \hline
 IH-$0.59$-$-90^\circ$  & $-++$ & $83.63$ & $36.02$ \\
 \hline
  IH-$0.59$-$+90^\circ$  & $-+-$ & $63.85$ & $45.08$ \\
 \hline
 IH-$0.41$-$-90^\circ$  & $--+$ & $72.55$ & $29.83$ \\
 
 \hline
 IH-$0.41$-$+90^\circ$  & $---$ & $52.77$ & $38.89$ \\
 
 \hline
\end{tabular}
\end{center}
 \caption{Expected $\nu_e$ and $\bar{\nu}_e$ appearance events of \nova for $1.36\times10^{21}$ ($1.25\times 10^{21}$) POT in $\nu$ ($\bar{\nu}$) mode and for ten different combinations of the unknown parameter values. The observed numbers of $\nu_e$ and $\bar{\nu}_e$ events are 82 and 33 respectively.}
  \label{nova_events}
\end{table}

\begin{table}
  \begin{center}
\begin{tabular}{|c|c|c|c|}
  \hline
  Hierarchy-$\sin^2\tz$-$\dcp$ & Label&
  $\nu_e$ Appearance events&$\bar{\nu}_e$ Appearance events\\
  
  \hline
   Vacuum-$0.5$-$0$ & $000$ & $77.64$ & $18.62$ \\
  \hline
  NH-$0.5$-$0$ & $+00$ & $82.84$& $17.88$\\
  \hline
  NH-$0.59$-$0$  & $++0$ & $94.07$ & $19.84$ \\
  \hline
 NH-$0.59$-$-90^\circ$ & $+++$ & $112.72$ & $17.55$ \\
 \hline
  NH-$0.59$-$+90^\circ$ & $++-$ & $77.85$ & $22.06$ \\
 \hline
 NH-$0.41$-$-90^\circ$  & $+-+$ & $91.85$ & $13.64$ \\
 \hline
 IH-$0.59$-$-90^\circ$  & $-++$ & $98.34$ & $19.12$ \\
 \hline
 IH-$0.59$-$+90^\circ$  & $-+-$ & $66.71$ & $24.36$ \\
 \hline
 IH-$0.41$-$-90^\circ$  & $--+$ & $81.32$ & $14.62$ \\
 
 \hline
 IH-$0.41$-$+90^\circ$  & $---$ & $49.68$ & $19.86$ \\
 
 \hline
\end{tabular}
\end{center}
 \caption{Expected $\nu_e$ and $\bar{\nu}_e$ appearance events of T2K for $1.97\times10^{21}$ ($1.63\times 10^{21}$) POT in $\nu$ ($\bar{\nu}$) mode and for ten different combinations of the unknown parameter values. The observed numbers of $\nu_e$ and $\bar{\nu}_e$ events are 107 and 15 respectively.}
  \label{t2k_events}
\end{table}
\section{Future prospects}
 \label{future}
\begin{figure}[H]
\centering
\includegraphics[width=1.0\textwidth]{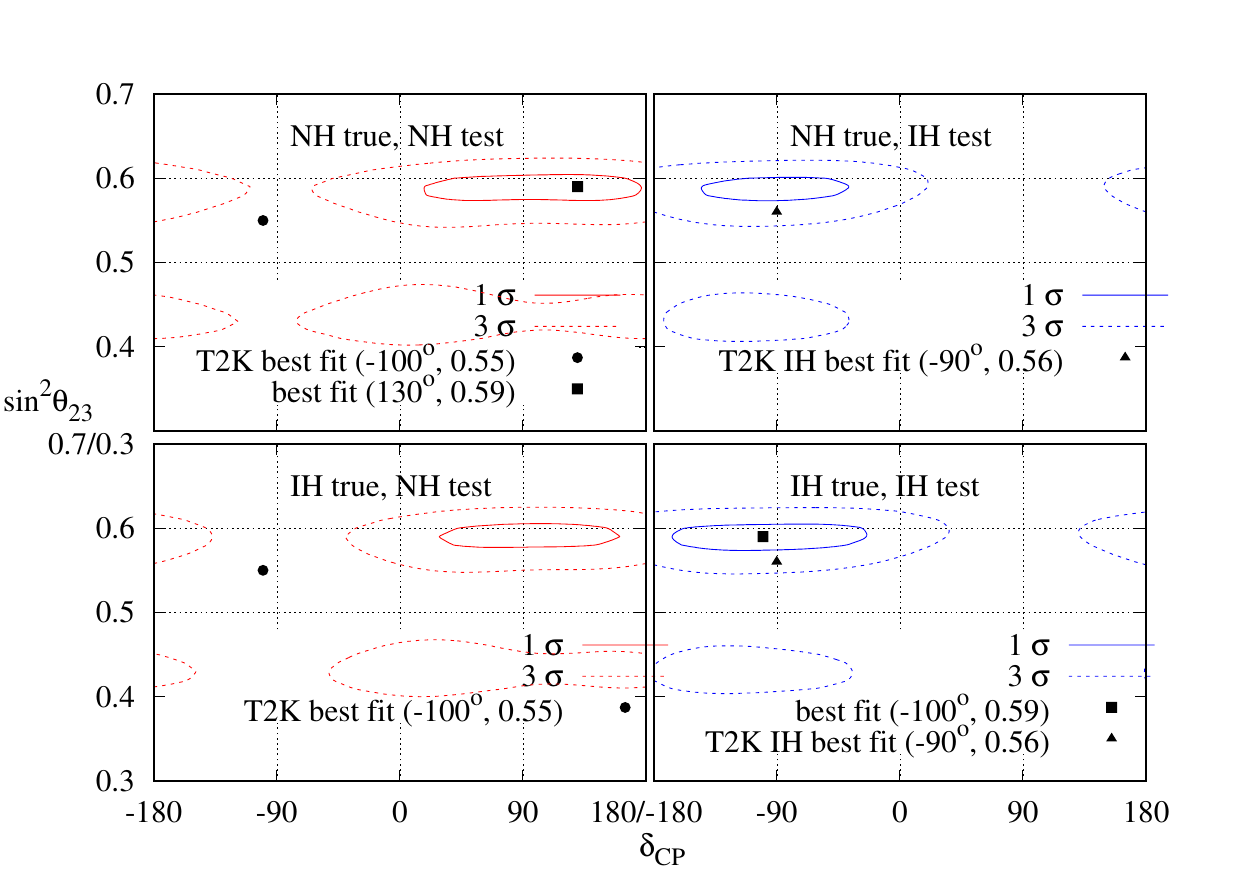}
\caption{\footnotesize{Allowed region in the $\sin^2 \tz-\dcp$ plane after equal $\nu$ and $\bar{\nu}$ run of total $30.25\times 10^{20}$ POT each for NO$\nu$A. True oscillation parameter values are fixed at current best fit values of \nova.
The upper (lower) panel shows the allowed regions for NH (IH) as true hierarchy. The left (right) panel represents test hierarchy NH (IH). The solid (dashed) lines indicate the allowed regions for $1\, \sigma$ ($3\, \sigma$) C.L.  
}}
\label{nova-5+5}
\end{figure}

If the tension between these two experiments is indeed due to a physical effect, it will be imperative to plan new experimental strategies to probe the nature of this new physics. In this section, we will discuss the possibility of establishing the tension between the two long baseline accelerator neutrino experiments with higher statistical significance in future. To do so, we have first simulated the \nova data for 5 years of neutrino and 5 years of anti-neutrino run, each corresponding to $30.25\times 10^{20}$ POT. We used the true parameter values as the current \nova best-fit values i.e. $\sin^2\tz=0.59$, $\dcp=130^{\circ}$ ($-100^{\circ}$), $\sin^2 2\ty=0.084$, $|\Delta m^{2}_{\rm eff}|=2.44\times10^{-3}\, {\rm eV}^2$, $\ds=7.39\times 10^{-5}\, {\rm eV}^2$ and $\sin^2\tx=0.31$ for NH (IH). We analysed these simulated data as described in section \ref{analysis}. Priors have been added on $\sin^2 2\ty$ only. The result has been presented on the $\sin^2\tz-\dcp$ plane in Fig.~\ref{nova-5+5}. If the current trend continues then \nova will be able to exclude the present T2K best fit point at $3\, \sigma$ C.L. for both NH and IH being the true hierarchy. T2K IH best-fit will be excluded only at $1\, \sigma$, for both NH and IH being the true hierarchy. We want to emphasize that when NH is the true hierarchy, \nova cannot exclude IH test at $1\, \sigma$ C.L. This is because of the hierarchy-$\dcp$ degeneracy. The combination of NH and $\dcp$ in UHP is unfavourable for determining the hierarchy and can be mimicked by the combination of IH and $\dcp$ in LHP due to the said degeneracy. Thus IH cannot be ruled out at $1\, \sigma$ for NH and $\dcp=130^\circ$. Similarly, when IH is the true hierarchy and true $\dcp=-100^\circ$, there is a $1\, \sigma$ allowed region at NH test and $\dcp$ test in UHP. The tension for NH test hierarchy, can be established at $3\, \sigma$ by \nova irrespective of the true hierarchy. It is to be noted that after 5 years each of $\nu$ and $\anu$ run, \nova can exclude most of the present T2K allowed region for NH (presented in fig. \ref{nu+nubar-app+disapp}) at $3\,\sigma$ C.L. for both the true hierarchies.

\begin{figure}[H]
\centering
\includegraphics[width=1.0\textwidth]{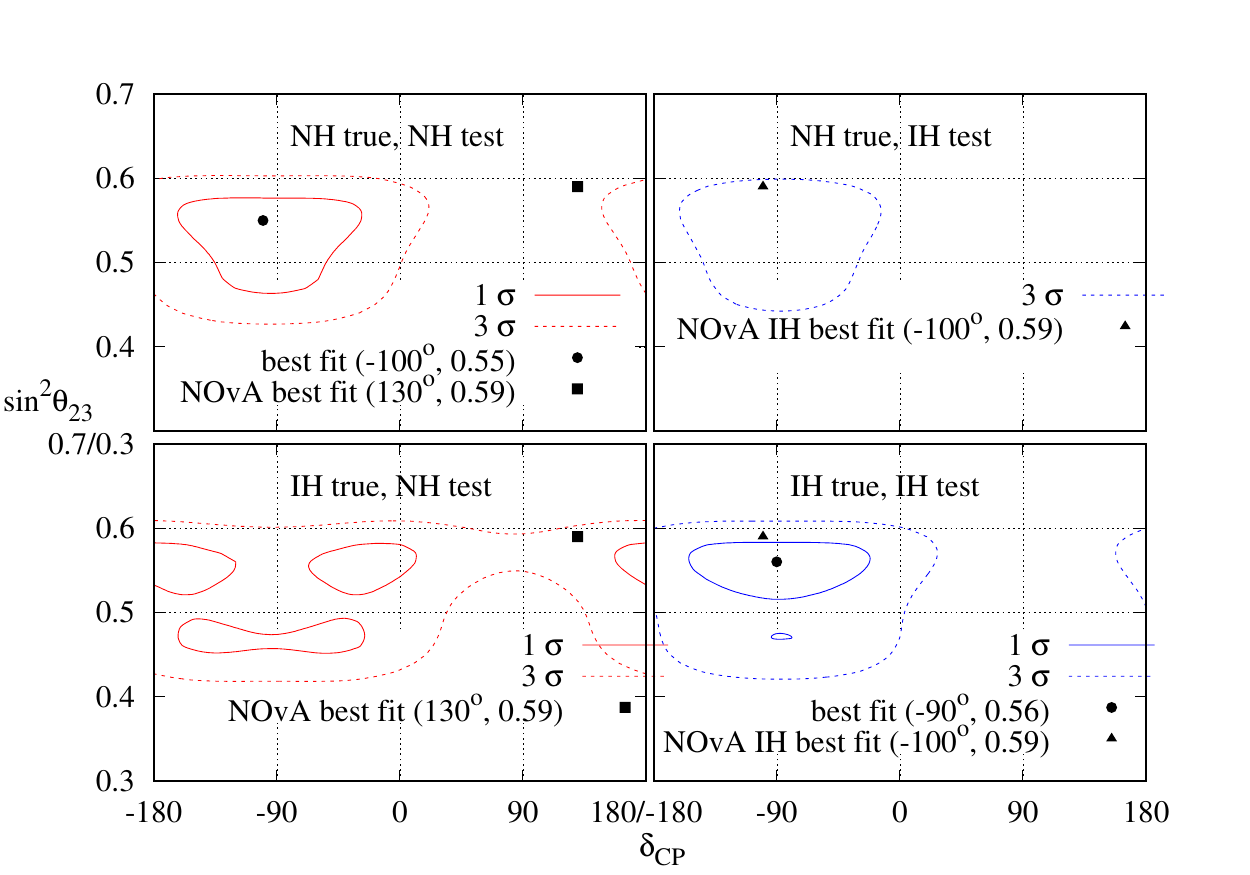}
\caption{\footnotesize{Allowed region in the $\sin^2 \tz-\dcp$ plane after equal $\nu$ and $\bar{\nu}$ run of total $37.4\times 10^{20}$ POT each for T2K. True oscillation parameter values are fixed at current best fit values of T2K.
The upper (lower) panel shows the allowed regions for NH (IH) as true hierarchy. The left (right) panel represents test hierarchy NH (IH). The solid (dashed) lines indicate the allowed regions for $1\, \sigma$ ($3\, \sigma$) C.L.  
}}
\label{T2K-5+5}
\end{figure}

In Fig.~\ref{T2K-5+5}, we have shown the expectation of T2K with equal 5 years of $\nu$ and 5 years of $\bar{\nu}$ run of total $37.4\times 10^{20}$ POT each. To do so, we simulated the experimental events with the present T2K best fit points, i.e. $\sin^2\tz=0.55$ ($0.56$), $\dcp=-100^{\circ}$ ($-90^\circ$), $\sin^2 2\ty=0.084$, $|\Delta m^{2}_{\rm eff}|=2.44\times10^{-3}\, {\rm eV}^2$, $\ds=7.39\times 10^{-5}\, {\rm eV}^2$ and $\sin^2\tx=0.31$ for NH(IH). The theoretical event rates were calculated in the same way described in the beginning. Prior has been added on $\sin^2 2\ty$ only. If the current trend continues then T2K will be able to exclude the present \nova best-fit point at $3\, \sigma$ C.L. if NH is the true hierarchy. Also, when NH is true, there would be no IH allowed region at $1\, \sigma$ C.L. This is because NH and $\dcp$ in LHP is favourable hierarchy-$\dcp$ combination to determine the hierarchy. However, because of a shorter baseline, matter effect is smaller in T2K, and thus IH test cannot be ruled out at $3\, \sigma$, even though the true hierarchy and $\dcp$ are in favourable combination. Since the matter effect is small, when IH is the true hierarchy and true $\dcp=-90^\circ$, \nova NH best-fit point can be ruled out at $1\,\sigma$ despite the hierarchy-$\dcp$ degeneracy. The said degeneracy, however, does not let the \nova NH best fit point to be ruled out at $3\, \sigma$ when IH is the true hierarchy and true $\dcp=-90^\circ$. The tension at NH test can be established by T2K at $3\, \sigma$ ($1\, \sigma$) if NH (IH) is the true hierarchy. After 5 years each of $\nu$ and $\anu$ run, T2K can exclude most of the present \nova allowed region for NH (presented in fig. \ref{nu+nubar-app+disapp}) at $3\,\sigma$ C.L. for both the true hierarchies.

The future long baseline accelerator neutrino oscillation experiment DUNE \cite{Abi:2018alz, Abi:2018dnh, Abi:2018rgm} consists of a baseline of $1300$ km. It has been designed to disentangle the changes due to CP from the changes due to matter effect. Its baseline and correspondingly its energy are much longer compared to \nova and T2K and hence the matter effect is much larger. Therefore, it is expected to measure the unknown quantities with much better precision. 

\begin{figure}[H]
\centering
\includegraphics[width=1.0\textwidth]{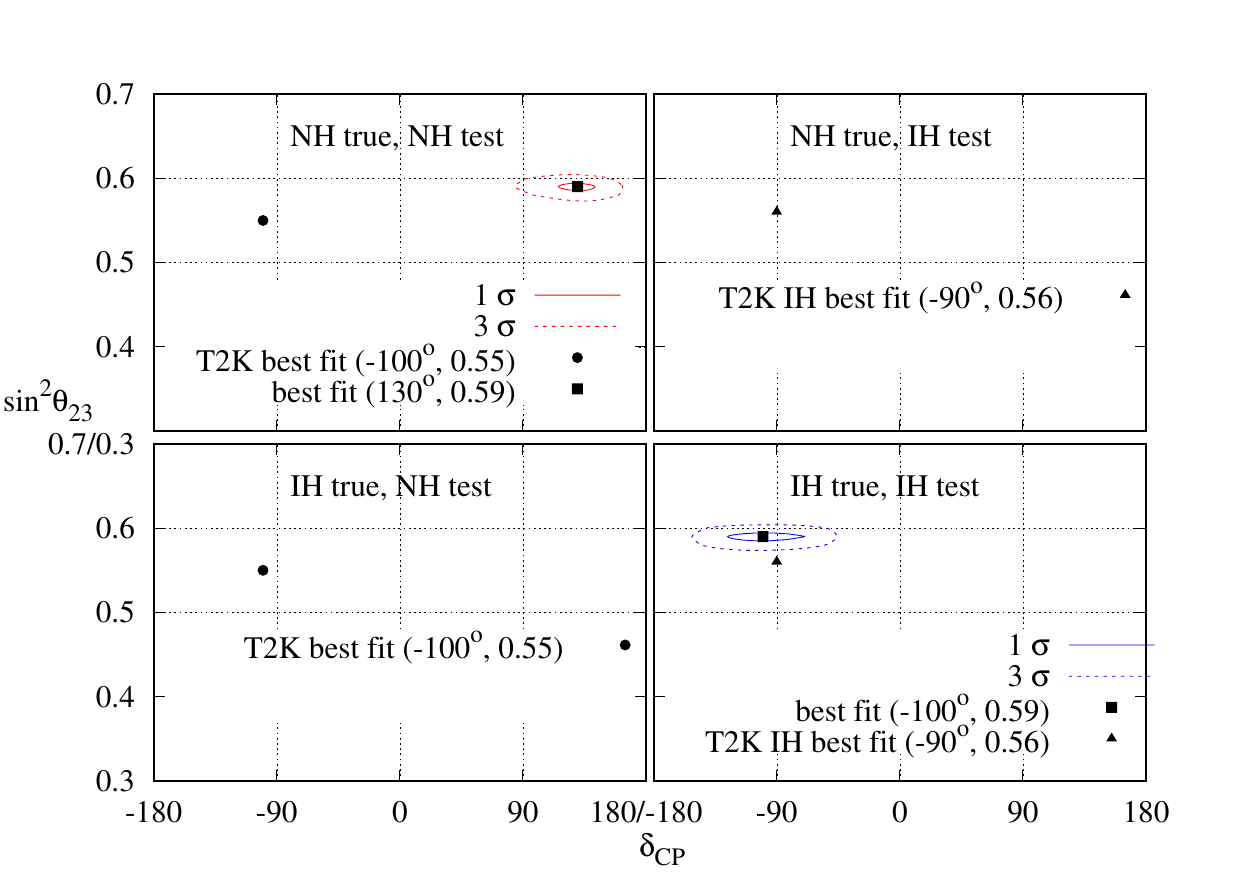}
\caption{\footnotesize{Allowed region in the $\sin^2 \tz-\dcp$ plane after equal $\nu$ and $\bar{\nu}$ run of total $73.5\times 10^{20}$ POT each for DUNE. True oscillation parameter values are fixed at current best fit values of \nova.
The upper (lower) panel shows the allowed regions for NH (IH) as true hierarchy. The left (right) panel represents test hierarchy NH (IH). The solid (dashed) lines indicate the allowed regions for $1\, \sigma$ ($3\, \sigma$) C.L.  
}}
\label{dune-5+5-nova}
\end{figure}

At first we have simulated the DUNE data for 5 years each of neutrino and anti-neutrino run, corresponding to $73.5\times 10^{20}$ POT each, with the true values as present \nova best-fit values. The analysis of this simulated data have been done as before with the priors on $\sin^22\ty$. From Fig.~\ref{dune-5+5-nova}, it can be seen that if the current trend of \nova continues, DUNE can rule out T2K best-fit points at $3\, \sigma$ for both NH and IH being the true hierarchy. Also, the wrong hierarchy is ruled out at $3\, \sigma$ for both the true hierarchy. It is to be noted that after 5 years each of $\nu$ and $\anu$ run, DUNE can exclude all of the present T2K allowed region for NH (presented in fig. \ref{nu+nubar-app+disapp}) at $3\,\sigma$ C.L. for both the true hierarchies if the true oscillation parameter values are the present \nova best-fit values.

\begin{figure}[H]
\centering
\includegraphics[width=1.0\textwidth]{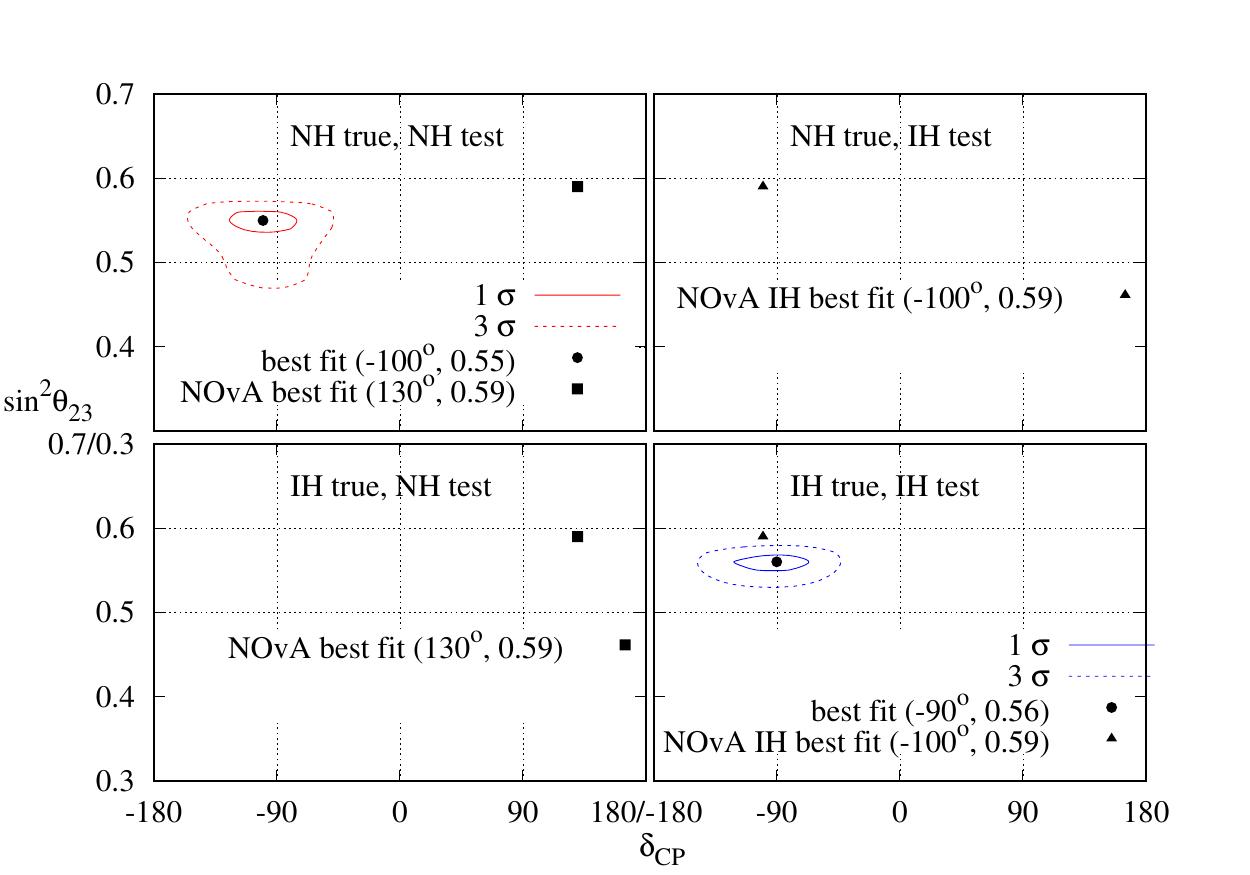}
\caption{\footnotesize{Allowed region in the $\sin^2 \tz-\dcp$ plane after equal $\nu$ and $\bar{\nu}$ run of total $73.5\times 10^{20}$ POT each for DUNE. True oscillation parameter values are fixed at current best fit values of T2K.
The upper (lower) panel shows the allowed regions for NH (IH) as true hierarchy. The left (right) panel represents test hierarchy NH (IH). The solid (dashed) lines indicate the allowed regions for $1\, \sigma$ ($3\, \sigma$) C.L.  
}}
\label{dune-5+5-t2k}
\end{figure}

After this, we have simulated the DUNE data for equal 5 years of neutrino and anti-neutrino run with the true values being equal to the T2K best-fit values. If the current T2K trend continues, then from Fig.~\ref{dune-5+5-t2k}, we can infer that \nova best-fit points can be ruled out at $3\, \sigma$ for both NH and IH being the true hierarchy. The wrong hierarchy can also be ruled out for both of the true hierarchies. After 5 years each of $\nu$ and $\anu$ run, DUNE can exclude all of the present \nova allowed region for NH (presented in fig. \ref{nu+nubar-app+disapp}) at $3\,\sigma$ C.L. for both the true hierarchies if the true oscillation parameter values are the present T2K best-fit values.

\section{Conclusion}
\label{conclusion}
From the analysis of the four present data sets from each of the two experiments \nova and T2K, we can conclude that there is not much discrepancy between the $\nu_\mu$ and $\bar{\nu}_\mu$ disappearance data of the two experiments. However, the $\bar{\nu}_\mu$ disappearance data of \nova rule out $\sin^2\tz\approx 0.5$ at $1\, \sigma$, whereas the same data from T2K allow the whole range of $\sin^2\tz$ at $1\, \sigma$. Therefore \nova $\bar{\nu}_\mu$ disappearance data have a mild conflict with all other disappearance data. The tension arises primarily from the $\nu_e$ appearance data of the two experiments, and is compounded by their $\nu_\mu$ disappearance data. Because of the very large number of observed $\nu_e$ appearance events in the T2K experiment, $\dcp$ anchors around $\dcp=-90^\circ$. All other appearance data sets, namely $\nu_e$ appearance from NO$\nu$A, $\bar{\nu}_e$ appearance from both \nova and T2K allow other values of $\dcp$ values different from $\dcp \approx -90^\circ$. The observed $\nu_e$ appearance event number of \nova is moderately large such that it can be accommodated by either NH-$\tz$ in HO-$\dcp$ in UHP or IH-$\tz$ in HO-$\dcp$ in LHP. The moderately large electron appearance events in \nova and very large electron appearance events in T2K are the biggest sources of tension between the two experiments and the addition of $\bar{\nu}_\mu$ disappearance data makes this tension even stronger.

Vacuum oscillation gives as good fit to the data from both the experiments as matter effect. There is a large overlap between the allowed regions at $1\, \sigma$ C.L. in the $\sin^2\tz-\dcp$ plane.

If the current trend continues, then after 5 years of neutrino and anti-neutrino run each, \nova (T2K) can exclude the T2K (NO$\nu$A) NH best-fit point at $3\, \sigma$ for both NH and IH (only NH) being the true hierarchy. Thus after 5 years of equal neutrino and anti-neutrino run, the tension at NH test can be established at $3\, \sigma$ by \nova irrespective of the true hierarchy. On the other hand, after the same run time, T2K can establish the tension at NH at $3\, \sigma$ only when NH is the true hierarchy. For IH being the true hierarchy, T2K can establish the tension at NH test only at $1\, \sigma$ after 5 years of equal neutrino and anti-neutrino runs. For both the true hierarchies, the IH best-fit point of one experiment can be ruled out by the other only at $1\, \sigma$. When IH is the true hierarchy, T2K can rule out the \nova NH best-fit point only at $1\, \sigma$. 

If the present tension continues, then DUNE after its 5 years each of neutrino and anti-neutrino run, can rule out the T2K (NO$\nu$A) NH and IH best-fit points at $3\, \sigma$ for both NH and IH being the true hierarchy. In such a case, we will have to look at the possibility of systematic effects or non-standard physics as the source of this tension.

\section{Acknowledgement}
We thank Suprabh Prakash, Soebur Razzaque, and S. Uma Sankar for valuable discussions, comments and suggestions.

\bibliographystyle{apsrev}
\bibliography{referenceslist}

\begin{thebibliography}{62}
\expandafter\ifx\csname natexlab\endcsname\relax\def\natexlab#1{#1}\fi
\expandafter\ifx\csname bibnamefont\endcsname\relax
  \def\bibnamefont#1{#1}\fi
\expandafter\ifx\csname bibfnamefont\endcsname\relax
  \def\bibfnamefont#1{#1}\fi
\expandafter\ifx\csname citenamefont\endcsname\relax
  \def\citenamefont#1{#1}\fi
\expandafter\ifx\csname url\endcsname\relax
  \def\url#1{\texttt{#1}}\fi
\expandafter\ifx\csname urlprefix\endcsname\relax\def\urlprefix{URL }\fi
\providecommand{\bibinfo}[2]{#2}
\providecommand{\eprint}[2][]{\url{#2}}

\bibitem[{\citenamefont{Bahcall et~al.}(2004)\citenamefont{Bahcall,
  Gonzalez-Garcia, and Pena-Garay}}]{Bahcall:2004ut}
\bibinfo{author}{\bibfnamefont{J.~N.} \bibnamefont{Bahcall}},
  \bibinfo{author}{\bibfnamefont{M.~C.} \bibnamefont{Gonzalez-Garcia}},
  \bibnamefont{and}
  \bibinfo{author}{\bibfnamefont{C.}~\bibnamefont{Pena-Garay}},
  \bibinfo{journal}{JHEP} \textbf{\bibinfo{volume}{08}}, \bibinfo{pages}{016}
  (\bibinfo{year}{2004}), \eprint{hep-ph/0406294}.

\bibitem[{\citenamefont{Ahmad et~al.}(2002)}]{Ahmad:2002jz}
\bibinfo{author}{\bibfnamefont{Q.~R.} \bibnamefont{Ahmad}} \bibnamefont{et~al.}
  (\bibinfo{collaboration}{SNO}), \bibinfo{journal}{Phys. Rev. Lett.}
  \textbf{\bibinfo{volume}{89}}, \bibinfo{pages}{011301}
  (\bibinfo{year}{2002}), \eprint{nucl-ex/0204008}.

\bibitem[{\citenamefont{Fukuda et~al.}(1998)}]{Super-Kamiokande:1998qwk}
\bibinfo{author}{\bibfnamefont{Y.}~\bibnamefont{Fukuda}} \bibnamefont{et~al.}
  (\bibinfo{collaboration}{Super-Kamiokande}), \bibinfo{journal}{Phys. Rev.
  Lett.} \textbf{\bibinfo{volume}{81}}, \bibinfo{pages}{1158}
  (\bibinfo{year}{1998}), \bibinfo{note}{[Erratum: Phys.Rev.Lett. 81, 4279
  (1998)]}, \eprint{hep-ex/9805021}.

\bibitem[{\citenamefont{Abe et~al.}(2008)}]{KamLAND:2008dgz}
\bibinfo{author}{\bibfnamefont{S.}~\bibnamefont{Abe}} \bibnamefont{et~al.}
  (\bibinfo{collaboration}{KamLAND}), \bibinfo{journal}{Phys. Rev. Lett.}
  \textbf{\bibinfo{volume}{100}}, \bibinfo{pages}{221803}
  (\bibinfo{year}{2008}), \eprint{0801.4589}.

\bibitem[{\citenamefont{Nichol}(2012{\natexlab{a}})}]{Kyoto2012MINOS}
\bibinfo{author}{\bibfnamefont{R.}~\bibnamefont{Nichol}}
  (\bibinfo{collaboration}{MINOS}) (\bibinfo{year}{2012}{\natexlab{a}}),
  \bibinfo{note}{talk given at the Neutrino 2012 Conference, June 3-9, 2012,
  Kyoto, Japan, \url{http://neu2012.kek.jp/}}.

\bibitem[{\citenamefont{Nunokawa et~al.}(2005)\citenamefont{Nunokawa, Parke,
  and Zukanovich~Funchal}}]{Nunokawa:2005nx}
\bibinfo{author}{\bibfnamefont{H.}~\bibnamefont{Nunokawa}},
  \bibinfo{author}{\bibfnamefont{S.~J.} \bibnamefont{Parke}}, \bibnamefont{and}
  \bibinfo{author}{\bibfnamefont{R.}~\bibnamefont{Zukanovich~Funchal}},
  \bibinfo{journal}{Phys.Rev.} \textbf{\bibinfo{volume}{D72}},
  \bibinfo{pages}{013009} (\bibinfo{year}{2005}), \eprint{hep-ph/0503283}.

\bibitem[{\citenamefont{Raut}(2012)}]{Raut:2012dm}
\bibinfo{author}{\bibfnamefont{S.~K.} \bibnamefont{Raut}}
  (\bibinfo{year}{2012}), \eprint{1209.5658}.

\bibitem[{\citenamefont{An et~al.}(2012)}]{An:2012eh}
\bibinfo{author}{\bibfnamefont{F.}~\bibnamefont{An}} \bibnamefont{et~al.}
  (\bibinfo{collaboration}{DAYA-BAY}), \bibinfo{journal}{Phys.Rev.Lett.}
  \textbf{\bibinfo{volume}{108}}, \bibinfo{pages}{171803}
  (\bibinfo{year}{2012}), \eprint{1203.1669}.

\bibitem[{\citenamefont{Ahn et~al.}(2012)}]{Ahn:2012nd}
\bibinfo{author}{\bibfnamefont{J.}~\bibnamefont{Ahn}} \bibnamefont{et~al.}
  (\bibinfo{collaboration}{RENO}), \bibinfo{journal}{Phys.Rev.Lett.}
  \textbf{\bibinfo{volume}{108}}, \bibinfo{pages}{191802}
  (\bibinfo{year}{2012}), \eprint{1204.0626}.

\bibitem[{\citenamefont{Abe et~al.}(2012)}]{Abe:2011fz}
\bibinfo{author}{\bibfnamefont{Y.}~\bibnamefont{Abe}} \bibnamefont{et~al.}
  (\bibinfo{collaboration}{Double Chooz}), \bibinfo{journal}{Phys.Rev.Lett.}
  \textbf{\bibinfo{volume}{108}}, \bibinfo{pages}{131801}
  (\bibinfo{year}{2012}), \eprint{1112.6353}.

\bibitem[{\citenamefont{De~Salas et~al.}(2018)\citenamefont{De~Salas, Gariazzo,
  Mena, Ternes, and T\'ortola}}]{deSalas:2018bym}
\bibinfo{author}{\bibfnamefont{P.~F.} \bibnamefont{De~Salas}},
  \bibinfo{author}{\bibfnamefont{S.}~\bibnamefont{Gariazzo}},
  \bibinfo{author}{\bibfnamefont{O.}~\bibnamefont{Mena}},
  \bibinfo{author}{\bibfnamefont{C.~A.} \bibnamefont{Ternes}},
  \bibnamefont{and}
  \bibinfo{author}{\bibfnamefont{M.}~\bibnamefont{T\'ortola}},
  \bibinfo{journal}{Front. Astron. Space Sci.} \textbf{\bibinfo{volume}{5}},
  \bibinfo{pages}{36} (\bibinfo{year}{2018}), \eprint{1806.11051}.

\bibitem[{\citenamefont{Albright and Chen}(2006)}]{Albright:2006cw}
\bibinfo{author}{\bibfnamefont{C.~H.} \bibnamefont{Albright}} \bibnamefont{and}
  \bibinfo{author}{\bibfnamefont{M.-C.} \bibnamefont{Chen}},
  \bibinfo{journal}{Phys.Rev.} \textbf{\bibinfo{volume}{D74}},
  \bibinfo{pages}{113006} (\bibinfo{year}{2006}), \eprint{hep-ph/0608137}.

\bibitem[{\citenamefont{Dolinski et~al.}(2019)\citenamefont{Dolinski, Poon, and
  Rodejohann}}]{Dolinski:2019nrj}
\bibinfo{author}{\bibfnamefont{M.~J.} \bibnamefont{Dolinski}},
  \bibinfo{author}{\bibfnamefont{A.~W.~P.} \bibnamefont{Poon}},
  \bibnamefont{and}
  \bibinfo{author}{\bibfnamefont{W.}~\bibnamefont{Rodejohann}},
  \bibinfo{journal}{Ann. Rev. Nucl. Part. Sci.} \textbf{\bibinfo{volume}{69}},
  \bibinfo{pages}{219} (\bibinfo{year}{2019}), \eprint{1902.04097}.

\bibitem[{\citenamefont{Hannestad}(2010)}]{Hannestad:2010kz}
\bibinfo{author}{\bibfnamefont{S.}~\bibnamefont{Hannestad}},
  \bibinfo{journal}{Prog. Part. Nucl. Phys.} \textbf{\bibinfo{volume}{65}},
  \bibinfo{pages}{185} (\bibinfo{year}{2010}), \eprint{1007.0658}.

\bibitem[{\citenamefont{Ayres et~al.}(2004)}]{Ayres:2004js}
\bibinfo{author}{\bibfnamefont{D.}~\bibnamefont{Ayres}} \bibnamefont{et~al.}
  (\bibinfo{collaboration}{NOvA}) (\bibinfo{year}{2004}),
  \eprint{hep-ex/0503053}.

\bibitem[{\citenamefont{Itow et~al.}(2001)}]{Itow:2001ee}
\bibinfo{author}{\bibfnamefont{Y.}~\bibnamefont{Itow}} \bibnamefont{et~al.}
  (\bibinfo{collaboration}{T2K}), pp. \bibinfo{pages}{239--248}
  (\bibinfo{year}{2001}), \eprint{hep-ex/0106019}.

\bibitem[{\citenamefont{Himmel}(2020)}]{Himmel:2020}
\bibinfo{author}{\bibfnamefont{A.}~\bibnamefont{Himmel}}
  (\bibinfo{year}{2020}), \bibinfo{note}{talk given at the Neutrino 2020
  meeting on July, 2nd,
  2020,\url{https://indico.fnal.gov/event/43209/contributions/187840/attachments/130740/159597/NOvA-Oscilations-NEUTRINO2020.pdf}}.

\bibitem[{\citenamefont{Acero et~al.}(2021)}]{NOvA:2021nfi}
\bibinfo{author}{\bibfnamefont{M.~A.} \bibnamefont{Acero}} \bibnamefont{et~al.}
  (\bibinfo{collaboration}{NOvA}) (\bibinfo{year}{2021}), \eprint{2108.08219}.

\bibitem[{\citenamefont{Dunne}(2020)}]{Dunne:2020}
\bibinfo{author}{\bibfnamefont{P.}~\bibnamefont{Dunne}} (\bibinfo{year}{2020}),
  \bibinfo{note}{talk given at the Neutrino 2020 meeting on July, 2nd,
  2020,\url{https://indico.fnal.gov/event/43209/contributions/187830/attachments/129636/159603/T2K_Neutrino2020.pdf}}.

\bibitem[{\citenamefont{Abe et~al.}(2021)}]{T2K:2021xwb}
\bibinfo{author}{\bibfnamefont{K.}~\bibnamefont{Abe}} \bibnamefont{et~al.}
  (\bibinfo{collaboration}{T2K}), \bibinfo{journal}{Phys. Rev. D}
  \textbf{\bibinfo{volume}{103}}, \bibinfo{pages}{112008}
  (\bibinfo{year}{2021}), \eprint{2101.03779}.

\bibitem[{\citenamefont{Kelly et~al.}(2021)\citenamefont{Kelly, Machado, Parke,
  Perez-Gonzalez, and Funchal}}]{Kelly:2020fkv}
\bibinfo{author}{\bibfnamefont{K.~J.} \bibnamefont{Kelly}},
  \bibinfo{author}{\bibfnamefont{P.~A.~N.} \bibnamefont{Machado}},
  \bibinfo{author}{\bibfnamefont{S.~J.} \bibnamefont{Parke}},
  \bibinfo{author}{\bibfnamefont{Y.~F.} \bibnamefont{Perez-Gonzalez}},
  \bibnamefont{and} \bibinfo{author}{\bibfnamefont{R.~Z.}
  \bibnamefont{Funchal}}, \bibinfo{journal}{Phys. Rev. D}
  \textbf{\bibinfo{volume}{103}}, \bibinfo{pages}{013004}
  (\bibinfo{year}{2021}), \eprint{2007.08526}.

\bibitem[{\citenamefont{Berns}(2021)}]{Berns:2021iss}
\bibinfo{author}{\bibfnamefont{L.}~\bibnamefont{Berns}}
  (\bibinfo{collaboration}{T2K}), in \emph{\bibinfo{booktitle}{{55th Rencontres
  de Moriond on Electroweak Interactions and Unified Theories}}}
  (\bibinfo{year}{2021}), \eprint{2105.06732}.

\bibitem[{\citenamefont{Miranda et~al.}(2021)\citenamefont{Miranda, Pasquini,
  Rahaman, and Razzaque}}]{Miranda:2019ynh}
\bibinfo{author}{\bibfnamefont{L.~S.} \bibnamefont{Miranda}},
  \bibinfo{author}{\bibfnamefont{P.}~\bibnamefont{Pasquini}},
  \bibinfo{author}{\bibfnamefont{U.}~\bibnamefont{Rahaman}}, \bibnamefont{and}
  \bibinfo{author}{\bibfnamefont{S.}~\bibnamefont{Razzaque}},
  \bibinfo{journal}{Eur. Phys. J. C} \textbf{\bibinfo{volume}{81}},
  \bibinfo{pages}{444} (\bibinfo{year}{2021}), \eprint{1911.09398}.

\bibitem[{\citenamefont{Chatterjee and Palazzo}(2021)}]{Chatterjee:2020kkm}
\bibinfo{author}{\bibfnamefont{S.~S.} \bibnamefont{Chatterjee}}
  \bibnamefont{and} \bibinfo{author}{\bibfnamefont{A.}~\bibnamefont{Palazzo}},
  \bibinfo{journal}{Phys. Rev. Lett.} \textbf{\bibinfo{volume}{126}},
  \bibinfo{pages}{051802} (\bibinfo{year}{2021}), \eprint{2008.04161}.

\bibitem[{\citenamefont{Denton et~al.}(2021)\citenamefont{Denton, Gehrlein, and
  Pestes}}]{Denton:2020uda}
\bibinfo{author}{\bibfnamefont{P.~B.} \bibnamefont{Denton}},
  \bibinfo{author}{\bibfnamefont{J.}~\bibnamefont{Gehrlein}}, \bibnamefont{and}
  \bibinfo{author}{\bibfnamefont{R.}~\bibnamefont{Pestes}},
  \bibinfo{journal}{Phys. Rev. Lett.} \textbf{\bibinfo{volume}{126}},
  \bibinfo{pages}{051801} (\bibinfo{year}{2021}), \eprint{2008.01110}.

\bibitem[{\citenamefont{Rahaman}(2021)}]{Rahaman:2021leu}
\bibinfo{author}{\bibfnamefont{U.}~\bibnamefont{Rahaman}},
  \bibinfo{journal}{Eur. Phys. J. C} \textbf{\bibinfo{volume}{81}},
  \bibinfo{pages}{792} (\bibinfo{year}{2021}), \eprint{2103.04576}.

\bibitem[{\citenamefont{Nizam et~al.}(2019)\citenamefont{Nizam, Bharti,
  Prakash, Rahaman, and Uma~Sankar}}]{Nizam:2018got}
\bibinfo{author}{\bibfnamefont{M.}~\bibnamefont{Nizam}},
  \bibinfo{author}{\bibfnamefont{S.}~\bibnamefont{Bharti}},
  \bibinfo{author}{\bibfnamefont{S.}~\bibnamefont{Prakash}},
  \bibinfo{author}{\bibfnamefont{U.}~\bibnamefont{Rahaman}}, \bibnamefont{and}
  \bibinfo{author}{\bibfnamefont{S.}~\bibnamefont{Uma~Sankar}},
  \bibinfo{journal}{Mod. Phys. Lett. A} \textbf{\bibinfo{volume}{35}},
  \bibinfo{pages}{06} (\bibinfo{year}{2019}), \eprint{1811.01210}.

\bibitem[{\citenamefont{Dohnal}(2021)}]{Dohnal:2021rcr}
\bibinfo{author}{\bibfnamefont{T.}~\bibnamefont{Dohnal}}
  (\bibinfo{collaboration}{Daya Bay}), \bibinfo{journal}{PoS}
  \textbf{\bibinfo{volume}{ICRC2021}}, \bibinfo{pages}{1175}
  (\bibinfo{year}{2021}).

\bibitem[{\citenamefont{Esteban et~al.}(2019)\citenamefont{Esteban,
  Gonzalez-Garcia, Hernandez-Cabezudo, Maltoni, and Schwetz}}]{Esteban:2018azc}
\bibinfo{author}{\bibfnamefont{I.}~\bibnamefont{Esteban}},
  \bibinfo{author}{\bibfnamefont{M.~C.} \bibnamefont{Gonzalez-Garcia}},
  \bibinfo{author}{\bibfnamefont{A.}~\bibnamefont{Hernandez-Cabezudo}},
  \bibinfo{author}{\bibfnamefont{M.}~\bibnamefont{Maltoni}}, \bibnamefont{and}
  \bibinfo{author}{\bibfnamefont{T.}~\bibnamefont{Schwetz}},
  \bibinfo{journal}{JHEP} \textbf{\bibinfo{volume}{01}}, \bibinfo{pages}{106}
  (\bibinfo{year}{2019}), \eprint{1811.05487}.

\bibitem[{\citenamefont{Nichol}(2012{\natexlab{b}})}]{Nichol:2012}
\bibinfo{author}{\bibfnamefont{R.}~\bibnamefont{Nichol}}
  (\bibinfo{collaboration}{MINOS}) (\bibinfo{year}{2012}{\natexlab{b}}),
  \bibinfo{note}{talk given at the Neutrino 2012 Conference, June 3-9, 2012,
  Kyoto, Japan, \url{http://neu2012.kek.jp/}}.

\bibitem[{\citenamefont{Huber et~al.}(2005)\citenamefont{Huber, Lindner, and
  Winter}}]{Huber:2004ka}
\bibinfo{author}{\bibfnamefont{P.}~\bibnamefont{Huber}},
  \bibinfo{author}{\bibfnamefont{M.}~\bibnamefont{Lindner}}, \bibnamefont{and}
  \bibinfo{author}{\bibfnamefont{W.}~\bibnamefont{Winter}},
  \bibinfo{journal}{Comput.Phys.Commun.} \textbf{\bibinfo{volume}{167}},
  \bibinfo{pages}{195} (\bibinfo{year}{2005}), \eprint{hep-ph/0407333}.

\bibitem[{\citenamefont{Huber et~al.}(2007)\citenamefont{Huber, Kopp, Lindner,
  Rolinec, and Winter}}]{Huber:2007ji}
\bibinfo{author}{\bibfnamefont{P.}~\bibnamefont{Huber}},
  \bibinfo{author}{\bibfnamefont{J.}~\bibnamefont{Kopp}},
  \bibinfo{author}{\bibfnamefont{M.}~\bibnamefont{Lindner}},
  \bibinfo{author}{\bibfnamefont{M.}~\bibnamefont{Rolinec}}, \bibnamefont{and}
  \bibinfo{author}{\bibfnamefont{W.}~\bibnamefont{Winter}},
  \bibinfo{journal}{Comput.Phys.Commun.} \textbf{\bibinfo{volume}{177}},
  \bibinfo{pages}{432} (\bibinfo{year}{2007}), \eprint{hep-ph/0701187}.

\bibitem[{\citenamefont{Acero et~al.}(2018)}]{NOvA:2018gge}
\bibinfo{author}{\bibfnamefont{M.~A.} \bibnamefont{Acero}} \bibnamefont{et~al.}
  (\bibinfo{collaboration}{NOvA}), \bibinfo{journal}{Phys. Rev. D}
  \textbf{\bibinfo{volume}{98}}, \bibinfo{pages}{032012}
  (\bibinfo{year}{2018}), \eprint{1806.00096}.

\bibitem[{\citenamefont{Acero et~al.}(2019)}]{NOvA:2019cyt}
\bibinfo{author}{\bibfnamefont{M.~A.} \bibnamefont{Acero}} \bibnamefont{et~al.}
  (\bibinfo{collaboration}{NOvA}), \bibinfo{journal}{Phys. Rev. Lett.}
  \textbf{\bibinfo{volume}{123}}, \bibinfo{pages}{151803}
  (\bibinfo{year}{2019}), \eprint{1906.04907}.

\bibitem[{nuf()}]{nufit}
\bibinfo{note}{\url{http://www.nu-fit.org/?q=node/45}}.

\bibitem[{\citenamefont{Wolfenstein}(1978)}]{msw1}
\bibinfo{author}{\bibfnamefont{L.}~\bibnamefont{Wolfenstein}},
  \bibinfo{journal}{Phys. Rev.} \textbf{\bibinfo{volume}{D17}},
  \bibinfo{pages}{2369} (\bibinfo{year}{1978}).

\bibitem[{\citenamefont{Mikheyev and Smirnov}(1985)}]{Mikheyev:1985zog}
\bibinfo{author}{\bibfnamefont{S.~P.} \bibnamefont{Mikheyev}} \bibnamefont{and}
  \bibinfo{author}{\bibfnamefont{A.~Y.} \bibnamefont{Smirnov}},
  \bibinfo{journal}{Sov. J. Nucl. Phys.} \textbf{\bibinfo{volume}{42}},
  \bibinfo{pages}{913} (\bibinfo{year}{1985}).

\bibitem[{\citenamefont{Mikheev and Smirnov}(1986)}]{Mikheev:1986wj}
\bibinfo{author}{\bibfnamefont{S.~P.} \bibnamefont{Mikheev}} \bibnamefont{and}
  \bibinfo{author}{\bibfnamefont{A.~Y.} \bibnamefont{Smirnov}},
  \bibinfo{journal}{Nuovo Cim. C} \textbf{\bibinfo{volume}{9}},
  \bibinfo{pages}{17} (\bibinfo{year}{1986}).

\bibitem[{\citenamefont{Fogli et~al.}(2006)\citenamefont{Fogli, Lisi, Marrone,
  and Palazzo}}]{Fogli:2005cq}
\bibinfo{author}{\bibfnamefont{G.~L.} \bibnamefont{Fogli}},
  \bibinfo{author}{\bibfnamefont{E.}~\bibnamefont{Lisi}},
  \bibinfo{author}{\bibfnamefont{A.}~\bibnamefont{Marrone}}, \bibnamefont{and}
  \bibinfo{author}{\bibfnamefont{A.}~\bibnamefont{Palazzo}},
  \bibinfo{journal}{Prog. Part. Nucl. Phys.} \textbf{\bibinfo{volume}{57}},
  \bibinfo{pages}{742} (\bibinfo{year}{2006}), \eprint{hep-ph/0506083}.

\bibitem[{\citenamefont{Bharti et~al.}(2021)\citenamefont{Bharti, Rahaman, and
  Uma~Sankar}}]{Bharti:2020gnu}
\bibinfo{author}{\bibfnamefont{S.}~\bibnamefont{Bharti}},
  \bibinfo{author}{\bibfnamefont{U.}~\bibnamefont{Rahaman}}, \bibnamefont{and}
  \bibinfo{author}{\bibfnamefont{S.}~\bibnamefont{Uma~Sankar}},
  \bibinfo{journal}{Mod. Phys. Lett. A} \textbf{\bibinfo{volume}{36}},
  \bibinfo{pages}{2150098} (\bibinfo{year}{2021}), \eprint{2001.08676}.

\bibitem[{\citenamefont{Abe et~al.}(2018)}]{Super-Kamiokande:2017yvm}
\bibinfo{author}{\bibfnamefont{K.}~\bibnamefont{Abe}} \bibnamefont{et~al.}
  (\bibinfo{collaboration}{Super-Kamiokande}), \bibinfo{journal}{Phys. Rev. D}
  \textbf{\bibinfo{volume}{97}}, \bibinfo{pages}{072001}
  (\bibinfo{year}{2018}), \eprint{1710.09126}.

\bibitem[{\citenamefont{Gandhi et~al.}(2007)\citenamefont{Gandhi, Ghoshal,
  Goswami, Mehta, Uma~Sankar, and Shalgar}}]{Gandhi:2007td}
\bibinfo{author}{\bibfnamefont{R.}~\bibnamefont{Gandhi}},
  \bibinfo{author}{\bibfnamefont{P.}~\bibnamefont{Ghoshal}},
  \bibinfo{author}{\bibfnamefont{S.}~\bibnamefont{Goswami}},
  \bibinfo{author}{\bibfnamefont{P.}~\bibnamefont{Mehta}},
  \bibinfo{author}{\bibfnamefont{S.}~\bibnamefont{Uma~Sankar}},
  \bibnamefont{and} \bibinfo{author}{\bibfnamefont{S.}~\bibnamefont{Shalgar}},
  \bibinfo{journal}{Phys. Rev. D} \textbf{\bibinfo{volume}{76}},
  \bibinfo{pages}{073012} (\bibinfo{year}{2007}), \eprint{0707.1723}.

\bibitem[{\citenamefont{Lipari}(2000)}]{Lipari:1999wy}
\bibinfo{author}{\bibfnamefont{P.}~\bibnamefont{Lipari}},
  \bibinfo{journal}{Phys. Rev. D} \textbf{\bibinfo{volume}{61}},
  \bibinfo{pages}{113004} (\bibinfo{year}{2000}), \eprint{hep-ph/9903481}.

\bibitem[{\citenamefont{Narayan and Sankar}(2000)}]{Narayan:1999ck}
\bibinfo{author}{\bibfnamefont{M.}~\bibnamefont{Narayan}} \bibnamefont{and}
  \bibinfo{author}{\bibfnamefont{S.~U.} \bibnamefont{Sankar}},
  \bibinfo{journal}{Phys. Rev. D} \textbf{\bibinfo{volume}{61}},
  \bibinfo{pages}{013003} (\bibinfo{year}{2000}), \eprint{hep-ph/9904302}.

\bibitem[{\citenamefont{Bharti et~al.}(2018)\citenamefont{Bharti, Prakash,
  Rahaman, and Uma~Sankar}}]{Bharti:2018eyj}
\bibinfo{author}{\bibfnamefont{S.}~\bibnamefont{Bharti}},
  \bibinfo{author}{\bibfnamefont{S.}~\bibnamefont{Prakash}},
  \bibinfo{author}{\bibfnamefont{U.}~\bibnamefont{Rahaman}}, \bibnamefont{and}
  \bibinfo{author}{\bibfnamefont{S.}~\bibnamefont{Uma~Sankar}},
  \bibinfo{journal}{JHEP} \textbf{\bibinfo{volume}{09}}, \bibinfo{pages}{036}
  (\bibinfo{year}{2018}), \eprint{1805.10182}.

\bibitem[{\citenamefont{Cervera et~al.}(2000)\citenamefont{Cervera, Donini,
  Gavela, Gomez~Cadenas, Hernandez, Mena, and Rigolin}}]{Cervera:2000kp}
\bibinfo{author}{\bibfnamefont{A.}~\bibnamefont{Cervera}},
  \bibinfo{author}{\bibfnamefont{A.}~\bibnamefont{Donini}},
  \bibinfo{author}{\bibfnamefont{M.~B.} \bibnamefont{Gavela}},
  \bibinfo{author}{\bibfnamefont{J.~J.} \bibnamefont{Gomez~Cadenas}},
  \bibinfo{author}{\bibfnamefont{P.}~\bibnamefont{Hernandez}},
  \bibinfo{author}{\bibfnamefont{O.}~\bibnamefont{Mena}}, \bibnamefont{and}
  \bibinfo{author}{\bibfnamefont{S.}~\bibnamefont{Rigolin}},
  \bibinfo{journal}{Nucl. Phys.} \textbf{\bibinfo{volume}{B579}},
  \bibinfo{pages}{17} (\bibinfo{year}{2000}), \bibinfo{note}{[Erratum: Nucl.
  Phys.B593,731(2001)]}, \eprint{hep-ph/0002108}.

\bibitem[{\citenamefont{Fogli and Lisi}(1996)}]{Fogli:1996pv}
\bibinfo{author}{\bibfnamefont{G.~L.} \bibnamefont{Fogli}} \bibnamefont{and}
  \bibinfo{author}{\bibfnamefont{E.}~\bibnamefont{Lisi}},
  \bibinfo{journal}{Phys.Rev.} \textbf{\bibinfo{volume}{D54}},
  \bibinfo{pages}{3667} (\bibinfo{year}{1996}), \eprint{hep-ph/9604415}.

\bibitem[{\citenamefont{Burguet-Castell
  et~al.}(2001)\citenamefont{Burguet-Castell, Gavela, Gomez-Cadenas, Hernandez,
  and Mena}}]{BurguetCastell:2001ez}
\bibinfo{author}{\bibfnamefont{J.}~\bibnamefont{Burguet-Castell}},
  \bibinfo{author}{\bibfnamefont{M.}~\bibnamefont{Gavela}},
  \bibinfo{author}{\bibfnamefont{J.}~\bibnamefont{Gomez-Cadenas}},
  \bibinfo{author}{\bibfnamefont{P.}~\bibnamefont{Hernandez}},
  \bibnamefont{and} \bibinfo{author}{\bibfnamefont{O.}~\bibnamefont{Mena}},
  \bibinfo{journal}{Nucl.Phys.} \textbf{\bibinfo{volume}{B608}},
  \bibinfo{pages}{301} (\bibinfo{year}{2001}), \eprint{hep-ph/0103258}.

\bibitem[{\citenamefont{Minakata and Nunokawa}(2001)}]{Minakata:2001qm}
\bibinfo{author}{\bibfnamefont{H.}~\bibnamefont{Minakata}} \bibnamefont{and}
  \bibinfo{author}{\bibfnamefont{H.}~\bibnamefont{Nunokawa}},
  \bibinfo{journal}{JHEP} \textbf{\bibinfo{volume}{0110}}, \bibinfo{pages}{001}
  (\bibinfo{year}{2001}), \eprint{hep-ph/0108085}.

\bibitem[{\citenamefont{Mena and Parke}(2004)}]{Mena:2004sa}
\bibinfo{author}{\bibfnamefont{O.}~\bibnamefont{Mena}} \bibnamefont{and}
  \bibinfo{author}{\bibfnamefont{S.~J.} \bibnamefont{Parke}},
  \bibinfo{journal}{Phys.Rev.} \textbf{\bibinfo{volume}{D70}},
  \bibinfo{pages}{093011} (\bibinfo{year}{2004}), \eprint{hep-ph/0408070}.

\bibitem[{\citenamefont{Prakash et~al.}(2012)\citenamefont{Prakash, Raut, and
  Sankar}}]{Prakash:2012az}
\bibinfo{author}{\bibfnamefont{S.}~\bibnamefont{Prakash}},
  \bibinfo{author}{\bibfnamefont{S.~K.} \bibnamefont{Raut}}, \bibnamefont{and}
  \bibinfo{author}{\bibfnamefont{S.~U.} \bibnamefont{Sankar}},
  \bibinfo{journal}{Phys.Rev.} \textbf{\bibinfo{volume}{D86}},
  \bibinfo{pages}{033012} (\bibinfo{year}{2012}), \eprint{1201.6485}.

\bibitem[{\citenamefont{Meloni}(2008)}]{Meloni:2008bd}
\bibinfo{author}{\bibfnamefont{D.}~\bibnamefont{Meloni}},
  \bibinfo{journal}{Phys.Lett.} \textbf{\bibinfo{volume}{B664}},
  \bibinfo{pages}{279} (\bibinfo{year}{2008}), \eprint{0802.0086}.

\bibitem[{\citenamefont{Agarwalla et~al.}(2013)\citenamefont{Agarwalla,
  Prakash, and Sankar}}]{Agarwalla:2013ju}
\bibinfo{author}{\bibfnamefont{S.~K.} \bibnamefont{Agarwalla}},
  \bibinfo{author}{\bibfnamefont{S.}~\bibnamefont{Prakash}}, \bibnamefont{and}
  \bibinfo{author}{\bibfnamefont{S.~U.} \bibnamefont{Sankar}}
  (\bibinfo{year}{2013}), \eprint{1301.2574}.

\bibitem[{\citenamefont{Nath et~al.}(2016)\citenamefont{Nath, Ghosh, and
  Goswami}}]{Nath:2015kjg}
\bibinfo{author}{\bibfnamefont{N.}~\bibnamefont{Nath}},
  \bibinfo{author}{\bibfnamefont{M.}~\bibnamefont{Ghosh}}, \bibnamefont{and}
  \bibinfo{author}{\bibfnamefont{S.}~\bibnamefont{Goswami}},
  \bibinfo{journal}{Nucl. Phys. B} \textbf{\bibinfo{volume}{913}},
  \bibinfo{pages}{381} (\bibinfo{year}{2016}), \eprint{1511.07496}.

\bibitem[{\citenamefont{Bora et~al.}(2016)\citenamefont{Bora, Ghosh, and
  Dutta}}]{Bora:2016tmb}
\bibinfo{author}{\bibfnamefont{K.}~\bibnamefont{Bora}},
  \bibinfo{author}{\bibfnamefont{G.}~\bibnamefont{Ghosh}}, \bibnamefont{and}
  \bibinfo{author}{\bibfnamefont{D.}~\bibnamefont{Dutta}},
  \bibinfo{journal}{Adv. High Energy Phys.} \textbf{\bibinfo{volume}{2016}},
  \bibinfo{pages}{9496758} (\bibinfo{year}{2016}), \eprint{1606.00554}.

\bibitem[{\citenamefont{Barger et~al.}(2002)\citenamefont{Barger, Marfatia, and
  Whisnant}}]{Barger:2001yr}
\bibinfo{author}{\bibfnamefont{V.}~\bibnamefont{Barger}},
  \bibinfo{author}{\bibfnamefont{D.}~\bibnamefont{Marfatia}}, \bibnamefont{and}
  \bibinfo{author}{\bibfnamefont{K.}~\bibnamefont{Whisnant}},
  \bibinfo{journal}{Phys. Rev. D} \textbf{\bibinfo{volume}{65}},
  \bibinfo{pages}{073023} (\bibinfo{year}{2002}), \eprint{hep-ph/0112119}.

\bibitem[{\citenamefont{Kajita et~al.}(2007)\citenamefont{Kajita, Minakata,
  Nakayama, and Nunokawa}}]{Kajita:2006bt}
\bibinfo{author}{\bibfnamefont{T.}~\bibnamefont{Kajita}},
  \bibinfo{author}{\bibfnamefont{H.}~\bibnamefont{Minakata}},
  \bibinfo{author}{\bibfnamefont{S.}~\bibnamefont{Nakayama}}, \bibnamefont{and}
  \bibinfo{author}{\bibfnamefont{H.}~\bibnamefont{Nunokawa}},
  \bibinfo{journal}{Phys.Rev.} \textbf{\bibinfo{volume}{D75}},
  \bibinfo{pages}{013006} (\bibinfo{year}{2007}), \eprint{hep-ph/0609286}.

\bibitem[{\citenamefont{Esteban et~al.}(2020)\citenamefont{Esteban,
  Gonzalez-Garcia, Maltoni, Schwetz, and Zhou}}]{Esteban:2020cvm}
\bibinfo{author}{\bibfnamefont{I.}~\bibnamefont{Esteban}},
  \bibinfo{author}{\bibfnamefont{M.~C.} \bibnamefont{Gonzalez-Garcia}},
  \bibinfo{author}{\bibfnamefont{M.}~\bibnamefont{Maltoni}},
  \bibinfo{author}{\bibfnamefont{T.}~\bibnamefont{Schwetz}}, \bibnamefont{and}
  \bibinfo{author}{\bibfnamefont{A.}~\bibnamefont{Zhou}},
  \bibinfo{journal}{JHEP} \textbf{\bibinfo{volume}{09}}, \bibinfo{pages}{178}
  (\bibinfo{year}{2020}), \eprint{2007.14792}.

\bibitem[{\citenamefont{Capozzi et~al.}(2021)\citenamefont{Capozzi,
  Di~Valentino, Lisi, Marrone, Melchiorri, and Palazzo}}]{Capozzi:2021fjo}
\bibinfo{author}{\bibfnamefont{F.}~\bibnamefont{Capozzi}},
  \bibinfo{author}{\bibfnamefont{E.}~\bibnamefont{Di~Valentino}},
  \bibinfo{author}{\bibfnamefont{E.}~\bibnamefont{Lisi}},
  \bibinfo{author}{\bibfnamefont{A.}~\bibnamefont{Marrone}},
  \bibinfo{author}{\bibfnamefont{A.}~\bibnamefont{Melchiorri}},
  \bibnamefont{and} \bibinfo{author}{\bibfnamefont{A.}~\bibnamefont{Palazzo}},
  \bibinfo{journal}{Phys. Rev. D} \textbf{\bibinfo{volume}{104}},
  \bibinfo{pages}{083031} (\bibinfo{year}{2021}), \eprint{2107.00532}.

\bibitem[{\citenamefont{Abi et~al.}(2018{\natexlab{a}})}]{Abi:2018alz}
\bibinfo{author}{\bibfnamefont{B.}~\bibnamefont{Abi}} \bibnamefont{et~al.}
  (\bibinfo{collaboration}{DUNE}) (\bibinfo{year}{2018}{\natexlab{a}}),
  \eprint{1807.10327}.

\bibitem[{\citenamefont{Abi et~al.}(2018{\natexlab{b}})}]{Abi:2018dnh}
\bibinfo{author}{\bibfnamefont{B.}~\bibnamefont{Abi}} \bibnamefont{et~al.}
  (\bibinfo{collaboration}{DUNE}) (\bibinfo{year}{2018}{\natexlab{b}}),
  \eprint{1807.10334}.

\bibitem[{\citenamefont{Abi et~al.}(2018{\natexlab{c}})}]{Abi:2018rgm}
\bibinfo{author}{\bibfnamefont{B.}~\bibnamefont{Abi}} \bibnamefont{et~al.}
  (\bibinfo{collaboration}{DUNE}) (\bibinfo{year}{2018}{\natexlab{c}}),
  \eprint{1807.10340}.

\end{thebibliography}


\end{document}